\documentclass[lettersize,journal]{IEEEtran}
\usepackage{amsmath,amsthm,amssymb,amscd}
\newtheorem{theorem}{Theorem}
\newtheorem{proposition}[theorem]{Proposition}
\newtheorem{assumption}[theorem]{Assumption}
\newtheorem{lemma}[theorem]{Lemma}
\newtheorem{corollary}[theorem]{Corollary}
\usepackage{mathtools}
\usepackage{array}
\usepackage{textcomp}
\usepackage{stfloats}
\usepackage{url}
\usepackage{verbatim}
\usepackage{graphicx}
\usepackage{cite}
\usepackage{xcolor}
\usepackage{cuted}
\usepackage{algorithm}
\usepackage{algpseudocode}
\usepackage{longtable}
\usepackage{booktabs}
\usepackage{balance}
\usepackage{diagbox}
\usepackage{stfloats}
\usepackage{subcaption} 
\usepackage{tikz}
\usepackage{preview}
\definecolor{blue}{rgb}{0,0,0}

\begin{document}


\title{Enhancing Vehicular Platooning with Wireless Federated Learning: A Resource-Aware Control Framework}

\author{Beining Wu,~\IEEEmembership{Member,~IEEE,} 
         Jun Huang,~\IEEEmembership{Senior Member,~IEEE,} 
         Qiang Duan,~\IEEEmembership{Senior Member,~IEEE,} 
        Liang (Leon) Dong,~\IEEEmembership{Senior Member,~IEEE,} and 
        Zhipeng Cai,~\IEEEmembership{Fellow,~IEEE}
\thanks{Beining Wu and Jun Huang are with the Department of Electrical Engineering and Computer Science, South Dakota State University, Brookings, 57006 SD. Email: Wu.Beining@jacks.sdstate.edu., Jun.Huang@sdstate.edu.}
\thanks{Qiang Duan is with the Department of Information Sciences and Technology, The Pennsylvania State University, Abington, 19001 PA. Email: qxd2@psu.edu.}
\thanks{Liang (Leon) Dong is with the Department of Electrical and Computer Engineering, Baylor University, Waco, 76798 TX. Email: Liang\_Dong@baylor.edu.}
\thanks{Zhipeng Cai is with the Department of Computer Science, Georgia State University, Atlanta, 30303 GA. E-mail: zcai@gsu.edu.}
\thanks{Manuscript received April 19, 2021; revised August 16, 2021.}}

\markboth{IEEE/ACM Transactions on Networking,~Vol.~14, No.~8, August~2021}%
{Shell \MakeLowercase{\textit{et al.}}: A Sample Article Using IEEEtran.cls for IEEE Journals}


\maketitle

\begin{abstract}

This paper aims to enhance the performance of Vehicular Platooning (VP) systems integrated with Wireless Federated Learning (WFL). In highly dynamic environments, vehicular platoons experience frequent communication changes and resource constraints, which significantly affect information exchange and learning model synchronization. To address these challenges, we first formulate WFL in VP as a joint optimization problem that simultaneously considers Age of Information (AoI) and Federated Learning Model Drift (FLMD) to ensure timely and accurate control. Through theoretical analysis, we examine the impact of FLMD on convergence performance and develop a two-stage \underline{R}esource-\underline{A}ware \underline{C}ontrol fram\underline{E}work (RACE). The first stage employs a Lagrangian dual decomposition method for resource configuration, while the second stage implements a multi-agent deep reinforcement learning approach for vehicle selection. The approach integrates Multi-Head Self-Attention and Long Short-Term Memory networks to capture spatiotemporal correlations in communication states. Experimental results demonstrate that, compared to baseline methods, the proposed framework improves AoI optimization by up to 45\%, accelerates learning convergence, and adapts more effectively to dynamic VP environments on the AI4MARS dataset.

\end{abstract}


\section{Introduction}
\subsection{Background}
Vehicular Platooning (VP) is an advanced paradigm in transportation technology that enables multiple vehicles coordinated by a platoon leader to travel closely together in a controlled fleet. VP leverages wireless communication systems, sensors, and autonomous driving technologies to maintain precise spacing and synchronized movements among the vehicles, significantly enhancing efficiency and safety \cite{Tan2024TITS,Li2023TFS}. Notably, VP has found applications in various areas. For example, VP can be used for tractor platooning in precision agriculture, enabling multiple tractors to work in tandem, reducing fuel consumption, and improving productivity. In military scenarios, VP systems navigate complex terrains and respond to rapidly changing environments, requiring robust perception and decision-making capabilities. Therefore, vehicular platooning is expected to revolutionize transportation and communication areas by enhancing operational efficiency and safety \cite{deere2024partnership,dot2024platooning}.

Machine Learning (ML) technologies have been employed in VP systems and played a crucial role in meeting application requirements. Conventional ML methods in VP require collecting and sharing vast amounts of data among the vehicles within the platoon. However, the need for continuous learning from diverse data sources on individual vehicles raises significant privacy concerns in VP systems. These privacy issues not only compromise sensitive information but also potentially affect system security, particularly when vehicles operate in sensitive environments or perform mission-critical tasks.

Wireless Federated Learning (WFL), which allows individual vehicles in a VP to collaboratively train a common ML model without exposing their raw data, offers a promising solution to this challenge. In a WFL-empowered VP system, the platoon leader first disseminates an initial model to individual vehicles, which use their local data to train the model and then upload the locally trained models to the leader for model aggregation. The aggregated global model is then downloaded to each vehicle for the next round of model training, and the iteration lasts till the model converges \cite{Pervej2025TWC, Huang2025TMC}. 
In addition to preserving privacy during model training, WFL optimizes the learning process by distributing the computational workload across the platoon and facilitates the efficient aggregation of local updates, securing the learning model's accuracy and responsiveness to dynamic conditions. As such, WFL-empowered VP may protect data privacy while achieving effective collaborative learning in dynamic environments.

Nevertheless, WFL-empowered VP faces unique challenges due to the highly dynamic and resource-constrained nature of vehicular wireless communications. The constantly evolving communication environment disrupts stable information exchange among vehicles, posing significant difficulties in maintaining coordination. For instance, adaptive speed adjustments within the platoon lead to rapid variations in relative positions, directly impacting inter-vehicle communication links and model synchronization. To sustain the learning performance in such a dynamic setting, WFL must ensure timely updates of ML models. Consequently, Age-of-Information (AoI), a metric quantifying the freshness of exchanged information, becomes a critical factor in facilitating real-time model updates \cite{Guo2024TON,Wu2025MNET}. On the other hand, deviations between locally trained models and the aggregated global model, referred to as Federated Learning Model Drift (FLMD), significantly affect both the security and efficiency of the learning process \cite{Sun2025IOTJ, Yang2025TNNLS, Dai2024TNNLS, Kang2024TII, Han2024IOTJ}. Addressing these challenges, \emph{this work aims to achieve timely and precise control in WFL-empowered VP systems to ensure coordinated and efficient operations in dynamic vehicular environments.}


\subsection{Motivation and Contributions}\

Recent advancements in WFL and VP systems have explored AoI optimization~\cite{FLAoI,FedAoI,Zhou2024IOTJ,Parvini2023TVT,Pervej2025TWC} and FLMD reduction~\cite{Yang2025TNNLS,Dai2024TNNLS,Kang2024TII,Han2024IOTJ} to enhance system performance. Researchers have proposed various resource-aware control strategies, including computational resource allocation \cite{Fan2023TITS,Zhu2024TON}, communication resource scheduling \cite{Wang2024INFOCOM,Pervej2024TWC}, client selection \cite{Yang2024TCCN,Ma2024TWC}, and task offloading \cite{Sun2023TMC,Liu2024TON}. Traditional convex optimization-based \cite{Fan2023TITS,FedAoI} and game theory-based \cite{Sun2023TMC} approaches offer theoretically optimal resource allocation solutions for static environments but lack adaptability in highly dynamic vehicular settings. In contrast, deep reinforcement learning (DRL)-based techniques leverage environment interaction and policy iteration to adapt to dynamic changes in communication topologies and system states \cite{MADRL1,Chen2024ICL,Zhao2023TVT,Parvini2023TVT}. While DRL improves system adaptability, existing methods predominantly optimize either AoI or FLMD individually, failing to jointly balance information timeliness and model consistency, which are inherently interdependent performance indicators. This limitation emphasizes the need for a holistic optimization framework that simultaneously addresses both factors to enhance the efficiency and reliability of WFL-empowered VP systems.

Despite the exciting progress made in the control of WFL-empowered VP systems, the following research gaps remain unaddressed in this important area.

\begin{enumerate}
    \item Existing studies primarily employ single-moment state representations or simplified sequence models, which cannot adequately capture long-term dependencies and spatiotemporal correlations in the system. This limitation in temporal modeling significantly hinders the effectiveness of control decisions, especially when predictive decisions based on historical communication states and model update patterns are needed.
    
    \item Existing studies predominantly focus on either optimizing AoI or improving FLMD, treating them as independent factors. However, in WFL-empowered VP systems, the interplay between AoI and FLMD directly affects system performance. Yet, current research lacks a cohesive optimization framework that simultaneously addresses both to achieve optimal communication efficiency and learning stability.
    
    \item Most existing WFL studies evaluate models on standard computer vision datasets such as MNIST or CIFAR-10, which fails to capture the complexities of real-world vehicular environments. In practical VP applications, vehicles may process terrain recognition, obstacle detection, and other tasks with high spatial heterogeneity and class imbalance. While some studies attempt to mitigate data heterogeneity, they lack validation on realistic datasets that simulate the diverse environmental conditions encountered in vehicular networks, limiting their practical applicability.
\end{enumerate}

To address these research gaps, this work designs a comprehensive framework for WFL-empowered VP systems that effectively captures temporal dynamics, jointly optimizes information freshness and model consistency, and validates performance on realistic heterogeneous datasets. Specifically, we make the following contributions.

\begin{itemize}

    \item We investigate a WFL-empowered VP system operating in a highly dynamic environment and formulate a joint optimization problem that simultaneously considers AoI and FLMD to ensure timely and accurate control. We characterize the impact of model drift on the convergence performance through a theoretical analysis of FLMD.

    \item We design a two-stage \underline{\textbf{R}}esource-\underline{\textbf{A}}ware \underline{\textbf{C}}ontrol fram\underline{\textbf{E}}work, termed \textbf{RACE}, to solve the formulated problem. In the first stage, we employ a Lagrangian dual decomposition method to optimize resource configuration, informing efficient allocation of computational and communication resources. In the second stage, we develop a multi-agent DRL approach for vehicle selection to enhance learning efficiency. To capture the spatial relationships between vehicles and the temporal evolution patterns of communication states, we establish an $M$-order Markov Decision Process (MDP) and design a Temporal Sequence Feature Extraction Network (TSFEN) that integrates Multi-Head Self-Attention (MHSA) and Long Short-Term Memory (LSTM) modules. With the theoretical analysis of FLMD, we also present an adaptive probability masking mechanism that dynamically adjusts device selection strategies. We show that our approach can improve learning efficiency while providing model security and robustness in highly dynamic conditions.

    \item We validate our approach using the AI4MARS dataset, which contains terrain annotations from the Curiosity, Opportunity, and Spirit Mars rovers. The dataset offers more realistic heterogeneity and class imbalance compared to commonly used datasets. The results demonstrate the effectiveness of our framework in handling non-IID data distributions in complex environments.
\end{itemize}

The remainder of this paper is structured as follows. Section \ref{RelatedWork} briefly summarizes the existing studies. Section \ref{SM} describes the system model. Section \ref{ProFm} formulates the resource allocation problem. Section \ref{Method} presents our resource-aware control framework. In Section \ref{PerfEvl}, we present and discuss simulation results. Section \ref{Conclusion} concludes this paper.

\section{Related Work} \label{RelatedWork}

\subsection{AoI-Oriented Optimization}

The optimization of AoI in WFL networks and VP systems has been extensively studied \cite{FLAoI, Ma2023ICC, Wang2024IOTJ, Parvini2023TVT, FedAoI, Zhou2024IOTJ, Wu2023access}. Ma {\sl et al.} proposed an algorithm leveraging the Lagrangian exponential method, which offers low complexity and scalability, to minimize both AoI and Channel State Information (CSI) uncertainty for each device. To further optimize AoI across the entire WFL framework and enhance federated learning update efficiency, the authors in \cite{FedAoI} employed the Karush-Kuhn-Tucker (KKT) conditions to decouple the AoI minimization problem. Their approach introduced virtual subchannels and utilized a matching-based algorithm to optimize AoI across multiple devices.

In the context of AoI optimization within VP systems, Zhou {\sl et al.} \cite{Zhou2024IOTJ} developed a joint communication and control model, leveraging queuing theory and stochastic geometry to analyze AoI distribution under dynamic vehicular conditions. Parvini {\sl et al.} \cite{Parvini2023TVT} extended this work by optimizing the transmission probabilities of AoI and Cooperative Awareness Messages (CAMs) within a multi-agent reinforcement learning (MARL) framework, utilizing the multi-agent deep deterministic policy gradient (MADDPG) algorithm. \textcolor{blue}{Recent advances have also explored Stackelberg game approaches for convergence acceleration~\cite{Wang2024TVT} and age-based scheduling policies for mobile edge networks~\cite{Yang2020ICASSP}, effectively addressing both communication efficiency and learning performance in dynamic environments.}

\subsection{Model Drift in Wireless Federated Learning}

The issue of model drift in WFL has gained increasing attention in recent studies \cite{Sun2025IOTJ, Yang2025TNNLS, Dai2024TNNLS, Kang2024TII, Han2024IOTJ}. Yang {\sl et al.} \cite{Yang2025TNNLS} introduced an allosteric feature collaboration method for model-heterogeneous federated learning, addressing the challenge that parameters of heterogeneous models cannot be directly aggregated. Their approach employs an allosteric feature generator to extract task-relevant information and facilitate knowledge exchange across clients, demonstrating effectiveness on standard FL benchmarks. Dai {\sl et al.} \cite{Dai2024TNNLS} proposed FedGAMMA, a global sharpness-aware minimization algorithm designed to mitigate client drift caused by non-IID data distributions. Unlike conventional methods that focus solely on local flatness, FedGAMMA aligns local updates with global flatness, significantly improving convergence stability and outperforming existing FL baselines.

To address both client drift and server drift under partial client participation, Kang {\sl et al.} \cite{Kang2024TII} developed FedAND, a unified optimization algorithm based on the consensus alternating direction method of multipliers (ADMM). Their method enhances convergence by suppressing server drift, which in turn reduces client drift, achieving more stable learning. From a security perspective, Han {\sl et al.} \cite{Han2024IOTJ} introduced FedDet, a robust algorithm designed to defend against adaptive model poisoning attacks. By splitting local models into layers for robust aggregation, their approach mitigates high-dimensionality challenges while preserving model functionality, demonstrating superior performance against adaptive adversarial attacks.

Despite these advancements in mitigating model consistency challenges, existing research lacks quantitative metrics to effectively measure model drift for resource allocation decisions. Furthermore, most studies do not account for the highly dynamic nature of vehicular environments, where communication conditions fluctuate and model consistency requirements evolve rapidly, limiting their applicability to WFL-empowered VP systems.

\subsection{Resource Management in VP}

Recent research on resource management in VP systems has primarily focused on both resource allocation and device selection strategies \cite{Fan2023TITS,Sun2023TMC,ChenICL2024,Hu2024TITS,Yang2024TCCN,WFLdelay2}. Fan {\sl et al.} \cite{Fan2023TITS} proposed a task offloading and resource allocation strategy based on convex optimization, aiming to minimize overall task processing delay for participating vehicles. However, its effectiveness is constrained by stringent mathematical assumptions, limiting adaptability in dynamic vehicular environments. To address VP dynamics, Hu {\sl et al.} \cite{Hu2024TITS} analyzed the impact of vehicular spacing adjustments on communication topologies, formulating a Markov Decision Process (MDP) with Multi-Agent Reinforcement Learning (MARL) to optimize power consumption. However, this approach fails to adequately address latency constraints and dynamic system state variations, which are critical for real-time federated learning.

In the context of device selection for WFL, Ouyang and Liu \cite{Yang2024TCCN} introduced a client selection framework based on contribution metrics to enhance learning efficiency, though it neglects energy constraints in resource-limited VP systems. Mao {\sl et al.} \cite{WFLdelay2} modeled device selection as an MDP and proposed a reinforcement learning framework to jointly minimize time costs and energy consumption. Despite these advancements, existing methods face critical limitations in WFL-empowered VP systems. Specifically, they predominantly prioritize network-level metrics over application-level learning performance, failing to capture the intricate relationship between communication efficiency and federated learning accuracy. Moreover, most approaches remain static in nature, making them ineffective in VP environments with continuously evolving communication topologies and dynamic resource constraints.

Unlike previous studies focusing solely on resource management in vehicular networks or federated learning optimization in static environments, our work optimizes AoI and FLMD in dynamic WFL-empowered VP systems. Existing approaches primarily emphasize metrics, such as latency or power consumption, without fully considering their impact on federated learning performance and model consistency. In contrast, our work integrates learning objectives and resource control in WFL-empowered VP systems by developing a two-stage optimization framework that combines Lagrangian dual decomposition for resource configuration with a multi-agent reinforcement learning-based vehicle selection strategy, enhanced by Multi-Head Self-Attention and Long Short-Term Memory mechanisms. Our approach effectively captures long-term dependencies and spatiotemporal correlations, securing adaptability to dynamic vehicular environments while improving both communication efficiency and learning stability.

\section{System Model} \label{SM}

\subsection{WFL-empowered Vehicle Platooning}

\begin{figure}[t]
    \centering
    \includegraphics[scale=0.6]{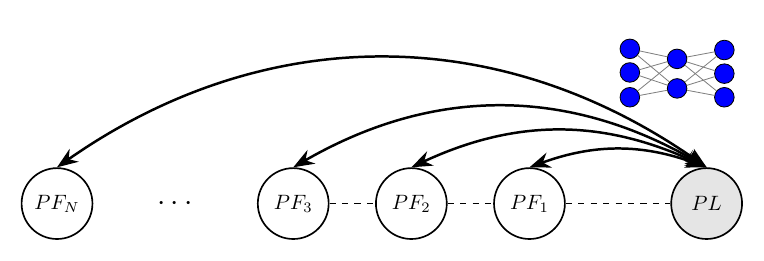}
    \caption{WFL-empowered Vehicle Platooning. The Platoon Leader (PL) coordinates with multiple Platoon Followers (PFs) through wireless communication links for model exchange.}
    \label{VP}
\end{figure}

We consider a WFL-empowered VP, which consists of a platoon leader (PL) and multiple platoon followers (PFs), as shown in Fig.\ref{VP}. In this configuration, the PL acts as the WFL server, coordinating a platoon with $N$ PFs and managing $K$ sub-channels, \textcolor{blue}{where $N \geq K$ due to practical vehicular communication limitations including spectrum regulatory constraints, interference management in dense environments, and computational capacity bounds for real-time federated learning}. The network operates using a Non-IID dataset, and each PF is equipped with a single antenna. The sets of all PFs and sub-channels are denoted as $\mathcal{N} = \{1, 2, \ldots, N\}$ and $\mathcal{K} = \{1, 2, \ldots, K\}$, respectively. The $n$-th PF in this WFL network can be represented as:
\begin{equation}
\mathcal{W}_n =\left\{\zeta_n, C_n\right\}, \forall n \in \mathcal{N},
\end{equation}
where $\zeta_n$ is the number of samples of the $n$-th PF, and $C_n$ is the configuration information in the communication process, including the channel status (assignment indicator), bandwidth, and power allocation.

\textcolor{blue}{The WFL-empowered VP learning process comprises four phases per communication round: 1) The PL broadcasts the global model to all PFs. 2) Each PF trains the model using its local dataset. 3) A subset of PFs is selected for aggregation due to limited sub-channels, and these PFs transmit their local models back to the PL. 4) The PL aggregates models and provides feedback for system optimization. We assume symmetric uplink/downlink conditions and negligible feedback transmission time due to significantly smaller instruction sizes compared to the model parameters.}


\begin{table}[t]
\centering
\renewcommand{\arraystretch}{1.3} 
\caption{Frequently used notations and symbols.}
\label{tab:key_notations}
\begin{tabular}{>{\arraybackslash}p{2.2cm} >{\arraybackslash}p{5.5cm}}
  \toprule
  \textbf{Notation} & \textbf{Definition} \\ 
  \midrule
  $N, K$  & Total number of PFs, the number of sub-channels \\
  $\mathcal{N}, \mathcal{S}[t]$ & Set of all PFs, subset of selected PFs at round $t$ \\
  $\omega[t], \omega_n[t]$ & The global model, the local model at round $t$ \\
  $\zeta_n, \xi$ & The number of samples of PF, the learning rate \\
  $\Theta_n[t], \Lambda_\Theta$ & The FLMD for the $n$-th PF, the FLMD threshold \\
  $x_n[t], \Delta x_n[t]$ & The position, the relative position \\
  $v_n[t], \Delta v_n[t]$ & The speed, the relative speed \\
  $\chi_n, \rho_n$ & The computational resource allocation index, the power allocation coefficient \\
  $G_n, \mu, \kappa$ & The available CPU cycles, required CPU cycles, power consumption coefficient \\
  $e_n^{\mathrm{cp}}, e_n^{\mathrm{tx}}$ & The computational energy consumption, the energy consumption for communication \\
  $\delta_n^{\mathrm{cp}}, \delta_n^{\mathrm{tx}}$ & The computational time, the transmission time \\
  $\delta_n, \delta[t]$ & The total time, the longest delay at round $t$ \\
  $B, \mathsf{P}$ &  The bandwidth of each sub-channel, the estimation error variance \\
  $|h_n|^2, D_n[t], P_n$ & The normalized channel gain, model size, maximum transmit power \\
  $\phi_{k,n}[t], r_n$ & The sub-channel assignment indicator, the achievable data rate \\
  $\alpha, \beta$ & Weighting coefficients for AoI and FLMD in the objective function \\
  $L, \eta$ & The Lipschitz smoothness coefficient, PL condition constant \\
  $\lambda_1, \lambda_2, \lambda_3, \lambda_4$ & Lagrangian multipliers for energy and power constraints \\
  $\Delta_n[t]$ & The AoI for the $n$-th PF at round $t$ \\
  \bottomrule
\end{tabular}
\end{table}

\subsection{Training Model}
In each communication round, the PL broadcasts the global model to all PFs. Each PF then trains the received model using its local dataset and subsequently transmits the updated local model back to the PL. Due to the limited availability of sub-channels, a subset $\mathcal{S}[t] \subseteq \mathcal{N}$ of PFs is selected for model aggregation in round $t$. The local loss and the global loss can be expressed as follows:
\begin{equation}
f_n\left(\omega[t]\right)=\frac{1}{\zeta_n} \sum_{i=1}^{\zeta_n} \ell\left(\omega[t] ; \mathbf{x}_{n, i}, y_{n, i}\right)
\end{equation}
and
\begin{equation}
\begin{aligned}
F\left(\omega[t], \mathcal{S}[t]\right) & =\sum_{n \in \mathcal{S}[t]} \frac{\zeta_n}{\sum_{n \in \mathcal{S}[t]} \zeta_n} f_n\left(\omega[t]\right) \\
& =\frac{\sum_{n \in \mathcal{S}[t]} \sum_{i=1}^{\zeta_n} \ell\left(\omega[t] ; \mathbf{x}_{n, i}, y_{n, i}\right)}{\sum_{n \in \mathcal{S}[t]} \zeta_n},
\end{aligned}
\end{equation}
where $\omega[t]$ is the global model in round $t$,  $\ell(\cdot)$ is a loss function, and $\left(\mathbf{x}_{n, i}, y_{n, i}\right)$ is the $i$-th sample of the $n$-th PF. 

The local model of the $n$-th PF in round $t$ can be expressed as:
\begin{equation}
\omega_n[t]=\omega[t]-\frac{\xi}{\zeta_n} \sum_{i=1}^{\zeta_n} \nabla \ell\left(\omega[t] ; \mathbf{x}_{n, i}, y_{n, i}\right),
\label{eq:local model}
\end{equation}
where $\xi$ is the learning rate. 

To quantify the deviation between the local models and the global model, we define the Federated Learning Model Drift (FLMD) for the $n$-th PF as:
\begin{equation}
\Theta_n[t] = \frac{\|\omega_n[t] - \omega[t]\|}{\|\omega[t]\|}.
\end{equation}

Here, $\|\cdot\|$ denotes the Euclidean norm, which measures the magnitude of a vector. Specifically, $\|\omega_n[t] - \omega[t]\|$ represents the Euclidean distance between the local model $\omega_n[t]$ and the global model $\omega[t]$. $\Theta_n[t]$ measures the relative drift of the local model from the global model, which reflects the impact of local data heterogeneity, communication delays, and resource constraints on training consistency.

From the perspective of security and privacy, $\Theta_n[t]$ serves as a screening metric to exclude anomalous devices that may compromise the global model due to poisoning attacks or abnormal behavior.

The server can update the global model based on federated averaging (FedAvg):
\begin{equation}
\omega[t+1]=\frac{\sum_{n \in \mathcal{S}[t]} \zeta_n \omega_n[t]}{\sum_{n \in \mathcal{S}[t]} \zeta_n}=\omega[t]-\xi \nabla F\left(\omega[t], \mathcal{S}[t]\right).
\label{eq:global model}
\end{equation}

To guarantee the integrity and robustness of the global model update, devices with $\Theta_n[t]$ exceeding a predefined threshold $\Lambda_{\Theta}$ are excluded from the aggregation set $\mathcal{E}[t]$:
\begin{equation}
\mathcal{E}[t] = \{n \in \mathcal{N} : \Theta_n[t] \leq \Lambda_{\Theta}\}.
\end{equation}

The above formula enhances the security of the WFL system by preventing devices with abnormal updates from compromising the learning process, while also preserving the privacy of legitimate devices by mitigating the risks of adversarial exploitation.

\subsection{Vehicle Platooning Model}

The position, speed, and acceleration of a vehicle (the $n$-th PF) in the platoon in round $t$ are denoted by $x_n[t], v_n[t], a_n[t]$. All vehicles move along a straight line, and the relative position and speed between devices are defined by the following relationships:
\begin{itemize}
    \item $\Delta x_n[t] = x_{n-1}[t] - x_n[t] - d_{n-1}$, where $d_{n-1}$ is the length of the preceding vehicle.
    \item $\Delta v_n[t] = v_n[t] - v_{n-1}[t]$, representing the relative speed.
\end{itemize}

The acceleration of the $n$-th PF can be calculated by the following equation:
\begin{equation}
\begin{split}
    a_n[t] &= \\
    &a_{\max }\left[1-\left(\frac{v_n[t]}{v_{\text {des }}}\right)^\delta-\left(\frac{H\left(v_n[t], \Delta v_n[t]\right)}{\Delta x_n[t]}\right)^2\right],
\end{split}
\end{equation}
where $a_{\max}$ is the maximum acceleration, $v_{\text{des}}$ is the desired velocity, and $\delta$ is the driver sensitivity feature, typically ranging from 1 to 5 \cite{RLVN}.

The function $H(\cdot)$ represents the safe distance between vehicles, calculated as:
\begin{equation}
H\left(v_n[t], \Delta v_n[t]\right)=d_{\min }+t_{\min } v_n[t]+\frac{v_n[t] \Delta v_n[t]}{2 \sqrt{a_{\max } b_{\max }}},
\end{equation}
where $d_{\min}$ is the minimum inter-vehicle space, $t_{\min}$ is the minimum reaction time and $b_{\max}$ is the maximum deceleration.

The speed and position of the PFs are updated as follows:
\begin{equation}
v_n[t]=v_n[t-1]+a_n[t-1] \cdot \tau, 
\end{equation}
and
\begin{equation}
x_n[t]=x_n[t-1]+v_n[t-1] \cdot \tau+\frac{1}{2} a_n[t-1] \cdot \tau^2,
\end{equation}
where $\tau$ is the update interval.

\subsection{Communication and Computation Models}

\subsubsection{Computation model} 

For each PF during the local training phase, the computational time is given by:
\begin{equation}
\delta_{n}^{\mathrm{cp}}={\mu \zeta_n}(\chi_{n} G_n)^{-1},
\end{equation}
where $\mu$ represents the required CPU cycles to train a single sample, the variable $\chi_{n} \in [0,1]$ denotes the computational resource allocation index for the $n$-th PF, and $G_n$ is the available CPU cycles at the $n$-th PF.

Accordingly, the computational energy consumption can be expressed as:
\begin{equation}
e_{n}^{\mathrm{cp}}=\kappa \mu \zeta_n\left(\chi_{n} G_n\right)^2,
\end{equation}
where $\kappa$ is the power consumption coefficient per CPU cycle.

\subsubsection{Communication model} 

During the communication stage, the local models are transmitted to the PL for aggregation. We define the size of the model for the $n$-th PF at the $t$-th communication round as $D_n[t]$. For the $n$-th PF, the achievable data rate is given by:
\begin{equation}
r_{n} = B \log_2\left(1 + \rho_{n} P_n \left|h_{n}\right|^2\right),
\label{eq:data_rate}
\end{equation}
where $B$ is the bandwidth of each sub-channel, $\rho_{ n} \in [0,1]$ is the power allocation coefficient, $P_n$ is the maximum transmit power of the $n$-th PF, and $\left|h_{n}\right|^2$ is the normalized channel gain of the $n$-th PF. Accurately measuring channel state information (CSI) by each transceiver in practical scenarios presents significant challenges. To mitigate the issue of imperfect CSI acquisition, we employ the minimum mean square error (MMSE) estimation technique~\cite{mmse}. Consequently, the composite channel gain can be formulated as
\begin{equation}
    \left|h_{n}\right|^2 = \sqrt{1- \mathsf{P}} \left|\hat{h}_{n}\right|^2 + \sqrt{\mathsf{P}} \left|\tilde{h}_{n}\right|^2, 
\end{equation}
where $\left|\hat{h}_{n}\right|^2 = \left|\hat{g}_n\right|^2 \eta d_n^{-\alpha}$ represents the estimate of channel gain, $\sigma^2$ is the noise variance, $\hat{g}_n \sim \mathcal{CN}(0,1)$ is the small-scale fading coefficient, $\eta$ is the frequency-dependent factor, $d_n[t] = x_0[t] - x_n[t]$ is the distance between the $n$-th PF and the PL, $\alpha$ is the path loss exponent, $\left|\tilde{h}_{n}\right|^2$ is the estimation error that is independent of $\left|\hat{h}_{n}\right|^2$, and parameter $\mathsf{P}$ represents the estimation error variance, taking a constant value from 0 to 1.

Under the condition of a constant data transmission rate during each communication round, the model transmission time for the $n$-th PF is:
\begin{equation}
    \delta_{n}^{\mathrm{tx}}=\frac{D_n[t]}{r_{ n}},
\end{equation}
and the energy consumption for communication is:
\begin{equation}
e_{n}^{\mathrm{tx}}=\rho_{n} P_n \delta_{ n}^{\mathrm{tx}}.
\end{equation}

\section{Problem Formulation and Theoretical Analysis} \label{ProFm}

\subsection{Problem Formulation}
Combining the previously established theoretical foundations, computational delay and communication delay, and the total time in any communication round for the $n$-th PF is:
\begin{equation}
\delta_{n}= \delta_{n}^{\mathrm{cp}}+ \delta_{n}^{\mathrm{tx}},
\end{equation}
and the time for the $t$-th communication round is determined by the PF with the longest delay, expressed as:
\begin{equation}
\delta{[t]}=\max _{n \in \mathcal{N}}\left\{\sum_{k \in \mathcal{K}} \phi_{k,n}{[t]} \delta_{n}\right\},
\end{equation}
where $\phi_{k, n}{[t]} \in\{0,1\}$ is the sub-channel assignment indicator. Specifically, $\phi_{k, n}{[t]}=1$ indicates that $n$-th PF is assigned to sub-channel $k$ in round $t$, while $\phi_{k, n}{[t]}=0$ indicates otherwise. During this communication round, the total energy consumption of $n$-th PF assigned to sub-channel $k$ is:
\begin{equation}
e_{n}=e_{n}^{\mathrm{cp}}+e_{n}^{\mathrm{tx}}.
\end{equation}

In practical scenarios, however, relying solely on the delay metric fails to comprehensively capture the timeliness and freshness of the information being communicated. These aspects are particularly crucial when real-time data is essential for decision-making processes. To address this limitation, we introduce the Age of Information (AoI) metric, which better reflects the timeliness of the transmitted data. AoI measures the time elapsed since the last received update was generated, thereby providing an effective characterization of the data’s freshness \cite{AoI2}.

While AoI reflects the timeliness of updates, it does not account for the quality of the information being communicated. Specifically, in WFL, the quality of updates is influenced by the deviation of local models from the global model, which we quantify using the FLMD. By considering both AoI and FLMD, our objective is to ensure not only timely but also high-quality updates for robust and secure global model training.

The inclusion of FLMD in the optimization process is motivated by the following considerations: (We will show the impact of FLMD on the convergence properties of the considered WFL system in the next subsection)
\begin{itemize}
    \item \emph{Model Consistency.} Minimizing FLMD ensures that the local models used for global aggregation remain consistent, thereby enhancing the stability and robustness of the global model.
    \item \emph{Security Screening.} Devices with excessive FLMD may indicate anomalies or adversarial behavior, and their exclusion improves the security of the WFL system.
    \item \emph{Privacy Protection.} Reducing FLMD mitigates the risk of exposing local data distributions during global aggregation.
\end{itemize}

For the $n$-th PF in the $t$-th round, if it is selected and establishes a channel, its AoI for that round is zero; otherwise:
\begin{equation}
\Delta_n{[t]} = \Delta_n{[t-1]}+\delta_{\max}{[t]},
\end{equation}
that is,
\begin{equation}
\Delta_n{[t]}=\left(1-\sum_{k=1}^K \phi_{k, n}{[t]}\right)\left(\Delta_n{[t-1]}+\delta_{\max}{[t]}\right).
\end{equation}

Thus, the optimization problem can be formulated as:
\begin{equation}
\min _{\phi, \chi, \rho} \sum_{t=1}^T \left( \lambda_1 \sum_{n=1}^N \Delta_n[t] + \lambda_2 \sum_{n=1}^N \Theta_n[t] \right)
\label{eq:AoI_FLMD}
\end{equation}
\begin{align*}
\text { s.t. } & \quad e_{n} \leq e_n^{\max },  \forall n \in \mathcal{N} \tag{\ref{eq:AoI_FLMD}a} \\
& \quad \chi_{n} \in[0,1],  \forall n \in \mathcal{N} \tag{\ref{eq:AoI_FLMD}b} \\
& \quad \rho_{n} \in[0,1],  \forall n \in \mathcal{N} \tag{\ref{eq:AoI_FLMD}c} \\
& \quad \phi_{k, n}{[t]} \in\{0,1\}, \forall k \in \mathcal{K}, \forall n \in \mathcal{N} \tag{\ref{eq:AoI_FLMD}d} \\
& \quad \sum_{n \in \mathcal{N}} \phi_{k, n}{[t]}=1, \forall k \in \mathcal{K} \tag{\ref{eq:AoI_FLMD}e} \\
& \quad \sum_{k \in \mathcal{K}} \phi_{k, n}{[t]} \in\{0,1\}, \forall n \in \mathcal{N} \tag{\ref{eq:AoI_FLMD}f} \\
& \quad \Theta_n[t] \leq \Lambda_{\Theta}, \forall n \in \mathcal{N}, \forall t \tag{\ref{eq:AoI_FLMD}g}
\end{align*}

The optimization variables in this problem are $\chi$, $\rho$, and $\phi$, which represent the resource allocation coefficient, power allocation coefficient, and channel allocation indicator, respectively. (\ref{eq:AoI_FLMD}a) imposes a constraint on the energy consumption for each communication round. (\ref{eq:AoI_FLMD}b)-(\ref{eq:AoI_FLMD}c) specify the permissible ranges for the resource allocation coefficient and power allocation coefficient. (\ref{eq:AoI_FLMD}d)-(\ref{eq:AoI_FLMD}f) ensure proper sub-channel allocation, and (\ref{eq:AoI_FLMD}g) limits the FLMD of each device to guarantee model robustness and security.

\subsection{Theoretical Analysis}
To analyze the impact of FLMD on the convergence properties of the considered WFL system, we establish the following theoretical foundation.

\begin{assumption}[Lipschitz Smooth Gradient]\label{assum:lipschitz}
The global loss function $F(\omega[t], \mathcal{N})$ is $L$-smooth, i.e.,
\begin{equation}
\|\nabla F(\omega[t-1], \mathcal{N})-\nabla F(\omega[t], \mathcal{N})\| 
\leq L\|\omega[t-1] - \omega[t]\|.
\end{equation}
\end{assumption}

\begin{assumption}[Polyak-Lojasiewicz Condition]\label{assum:pl}
The global loss function satisfies:
\begin{equation}
\|\nabla F(\omega[t], \mathcal{N})\|^2 \geq 2\eta[F(\omega[t], \mathcal{N}) - F(\omega^*, \mathcal{N})],
\end{equation}
where $\eta$ is a positive constant and $\omega^*$ is the optimal model.
\end{assumption}

\begin{lemma}\label{lemma:deviation}
Under Assumptions \ref{assum:lipschitz} and \ref{assum:pl}, for any device subset $\mathcal{S}[t] \subseteq \mathcal{N}$, the deviation term $e[t] = \nabla F(\omega[t], \mathcal{S}[t]) - \nabla F(\omega[t], \mathcal{N})$ satisfies:
\begin{multline}
\mathbb{E}[\|e[t]\|^2] \leq \\
\frac{(1-\frac{K}{N})\sum_{n\in\mathcal{N}}\zeta_n^2\|\nabla f_n(\omega[t])-\nabla F(\omega[t], \mathcal{N})\|^2}
{K(N-1)(\frac{1}{N}\sum_{n\in\mathcal{N}}\zeta_n)^2}.
\end{multline}
\end{lemma}

\begin{proof}
The detailed proof is provided in Appendix A.
\end{proof}



\begin{theorem}[FLMD Impact on Convergence]\label{thm:convergence}
Considering the impact of FLMD on convergence, the global loss reduction after $t$ communication rounds is bounded by:
\begin{multline}
\mathbb{E}[F(\omega[t + 1], \mathcal{N}) - F(\omega^*)] \leq \\
\left(1 - \frac{\mu}{L}\right)^t \mathbb{E}[F(\omega[1], \mathcal{N}) - F(\omega^*)] \\
+ \frac{1}{2L} \sum_{i=1}^{t} \left(1 - \frac{\mu}{L}\right)^{t-i} \mathbb{E}[\|e[i]\|^2], \label{eq:convergence_bound}
\end{multline}
where the learning rate satisfies $\xi = \frac{1}{L}$.
\end{theorem}

\textcolor{blue}{\begin{theorem}[Non-IID Sensitivity Analysis]\label{thm:noniid_sensitivity}
Under Assumptions \ref{assum:lipschitz} and \ref{assum:pl}, the convergence bound can be expressed in terms of data heterogeneity as:
\begin{multline}
\mathbb{E}[F(\omega[t+1], \mathcal{N}) - F(\omega^*)] \leq \\
\left(1 - \frac{\mu}{L}\right)^t\mathbb{E}[F(\omega[1], \mathcal{N}) - F(\omega^*)] \\
+ \frac{\gamma^2(1-\frac{K}{N})}{2LK(\bar{\zeta})^2}\sum_{i=1}^{t}\left(1 - \frac{\mu}{L}\right)^{t-i},
\end{multline}
where $\gamma^2 = \max_{n\in\mathcal{N}} \zeta_n^2\|\nabla f_n(\omega[t])-\nabla F(\omega[t], \mathcal{N})\|^2$ quantifies the worst-case local-global gradient divergence (data heterogeneity), and $\bar{\zeta} = \frac{1}{N}\sum_{n\in\mathcal{N}}\zeta_n$. 
\end{theorem}}

\begin{proof}
\textcolor{blue}{The detailed proof is provided in Appendix A.}
\end{proof}

\textcolor{blue}{\begin{corollary}[Tightness Conditions]\label{cor:tightness}
The bound in Theorem \ref{thm:noniid_sensitivity} is tight when: 1) $\gamma^2 \approx \mathbb{E}[\|\nabla f_n(\omega[t])-\nabla F(\omega[t], \mathcal{N})\|^2]$ (the worst-case gradient divergence approaches the expected case); 2) The selected subset $\mathcal{S}[t]$ contains devices with maximum gradient divergence; 3) $K \ll N$ (sparse participation regime).
\end{corollary}}

\textcolor{blue}{This analysis reveals that the convergence rate degrades proportionally to the data heterogeneity measure $\gamma^2$, providing quantitative sensitivity bounds for non-IID scenarios. The partial participation factor $(1-\frac{K}{N})$ demonstrates that limited vehicle selection can improve convergence by reducing gradient estimation variance when heterogeneous devices are excluded.}

\textcolor{blue}{\begin{theorem}[Adaptive Threshold Mechanism]\label{thm:adaptive_threshold}
To enhance device selection flexibility, we propose an adaptive threshold mechanism:
\begin{multline}
\Lambda_{\Theta}[t] = \Lambda_{\Theta}^{\min} + (\Lambda_{\Theta}^{\max} - \Lambda_{\Theta}^{\min}) \\
\cdot \exp\left(-\beta \frac{\|\nabla F(\omega[t], \mathcal{N})\|^2}{\|\nabla F(\omega[0], \mathcal{N})\|^2}\right),
\end{multline}
where $\Lambda_{\Theta}^{\min}$ and $\Lambda_{\Theta}^{\max}$ are the minimum and maximum allowable thresholds, and $\beta > 0$ controls the adaptation rate. Under this mechanism, the convergence bound becomes:
\begin{multline}
\mathbb{E}[F(\omega[t+1], \mathcal{N}) - F(\omega^*)] \leq \\
\left(1 - \frac{\mu}{L}\right)^t\mathbb{E}[F(\omega[1], \mathcal{N}) - F(\omega^*)] \\
+ \frac{1}{2L}\sum_{i=1}^{t}\left(1 - \frac{\mu}{L}\right)^{t-i}\mathbb{E}[\|e_{a}[i]\|^2],
\end{multline}
where $\mathbb{E}[\|e_{a}[i]\|^2] \leq \rho[i] \cdot \mathbb{E}[\|e[i]\|^2]$ with $\rho[i] = \frac{|\mathcal{M}_{a}[i]|}{|\mathcal{M}_{f}[i]|} \geq 1$. 
\end{theorem}}
\begin{proof}
\textcolor{blue}{The detailed proof is provided in Appendix A.}
\end{proof}

\textcolor{blue}{\begin{theorem}[Robustness under Relaxed Lipschitz Condition]\label{thm:relaxed_lipschitz}
When the global loss function is only locally $L$-smooth in a neighborhood $\mathcal{B}_R(\omega^*) = \{\omega: \|\omega - \omega^*\| \leq R\}$, our convergence analysis holds with probability $1-\delta$ if the initialization satisfies $\|\omega[0] - \omega^*\| \leq R - \epsilon$ for some $\epsilon > 0$, and the step size is chosen as $\xi \leq \frac{\epsilon}{2LT\sqrt{\log(1/\delta)}}$.
\end{theorem}}

\begin{proof}
\textcolor{blue}{The detailed proof is provided in Appendix A.}
\end{proof}

\textcolor{blue}{\begin{theorem}[Convergence without Polyak-Lojasiewicz]\label{thm:no_pl}
Without the PL condition, our algorithm still achieves convergence to stationary points. Specifically, the minimum gradient norm satisfies:
\begin{multline}
\min_{t=1,\cdots,T} \mathbb{E}[\|\nabla F(\omega[t], \mathcal{N})\|^2] \\
\leq \frac{2L(F(\omega[0]) - F^*) + \sum_{t=1}^T \mathbb{E}[\|e[t]\|^2]}{T},
\end{multline}
where $F^*$ is the global minimum value.
\end{theorem}}
\begin{proof}
\textcolor{blue}{The detailed proof is provided in Appendix A.}
\end{proof}

\textcolor{blue}{\begin{corollary}[Practical Robustness Guidelines]\label{cor:robustness}
For VP systems with non-convex terrain classification models, our framework maintains effectiveness when: 1) Models are initialized using pre-trained weights (ensuring proximity to good local minima); 2) FLMD thresholds are adapted based on local gradient norms rather than fixed values; 3) Learning rates are decreased progressively to handle local non-smoothness.
\end{corollary}}

\begin{theorem}\label{thm:flmd_bound}
For any PF $n$, its FLMD $\Theta_n[t]$ provides a direct measurement of local model deviation and satisfies:
\begin{equation}
\Theta_n[t] = \frac{\xi\|\nabla f_n(\omega[t])\|}{\|\omega[t]\|},
\end{equation}
with the upper bound:
\begin{equation}
\Theta_n[t] \leq \frac{\xi(L\|\omega[t] - \omega^*\| + \|\nabla f_n(\omega^*)\|)}{\|\omega[t]\|}.
\end{equation}
\end{theorem}

\begin{proof}
The detailed proof is provided in Appendix A.
\end{proof}





\section{Proposed Approach} \label{Method}

\subsection{RACE: A Resource-Aware Control Framework}

We design a two-stage \underline{\textbf{R}}esource-\underline{\textbf{A}}ware \underline{\textbf{C}}ontrol fram\underline{\textbf{E}}work (\textbf{RACE}), to address the optimization problem formulated in Section~\ref{ProFm}. As shown in \textbf{Algorithm~\ref{alg:race}}, The first stage employs a \emph{Lagrangian} dual decomposition method to determine the optimal computational and communication resource configuration for each PF. This stage provides each vehicle with the appropriate resources needed for efficient local model training and transmission, minimizing overall delay while enforcing energy constraints.


\begin{algorithm}[t]
\caption{RACE: Resource-Aware Control framEwork}
\label{alg:race}
\begin{algorithmic}[1]
\State \textbf{Stage 1:} Resource Configuration via Lagrangian Dual Decomposition
\For{each PF $n \in \mathcal{N}$}
    \State Determine optimal computational resource allocation coefficient; 
    \State Determine optimal power allocation coefficient;
    \State Configure PF's computation and communication resources accordingly;
\EndFor

\State \textbf{Stage 2:} Vehicle Selection via Multi-Agent Reinforcement Learning
\State Initialize learning agents for each sub-channel $k \in \mathcal{K}$
\For{each communication round $t = 1,2,\ldots,T$}
    \State Calculate FLMD for each PF;
    \For{each agent $k \in \mathcal{K}$}
        \State Observe current system state;
        \State Select vehicle for model aggregation based on MARL agents;
    \EndFor
    \State Aggregate models from selected vehicles;
    \State Update AoI and FLMD metrics;
\EndFor
\end{algorithmic}
\end{algorithm}

The second stage leverages multi-agent reinforcement learning  (MARL) to dynamically select vehicles for model aggregation in each communication round. By considering both AoI and FLMD during the selection process, this stage ensures that the global model is updated with both timely and accurate information from the most suitable vehicles.

By decomposing the optimization process into two distinct stages, RACE effectively addresses the formulated problem. Specifically, the resource configuration stage lays the foundation for efficient system operation by optimizing computational and communication resource allocation, while the vehicle selection stage ensures continuous adaptation and optimization as the platoon navigates dynamic environments. The following sections provide a detailed explanation of each stage and introduce the proposed approach in depth.

\subsection{Lagrangian Dual Decomposition for Optimal Resource Allocation}

By analyzing problem (\ref{eq:AoI_FLMD}), we decompose it into tractable sub-problems. For the $n$-th PF, the delay minimization problem is:
\begin{equation}
\min_{\chi_n,\rho_n} \frac{\mu \zeta_n}{\chi_n G_n} + \frac{D_n[t]}{B \log_2(1 + \rho_n P_n |h_n|^2)}
\label{eq:delay_min}
\end{equation}
\begin{equation}
\begin{split}
\text{s.t.} & \quad \kappa \mu \zeta_n (\chi_n G_n)^2 + \rho_n P_n \frac{D_n[t]}{B \log_2(1 + \rho_n P_n |h_n|^2)} \leq e_n^{\max}
\end{split}
\tag{\ref{eq:delay_min}a}
\end{equation}
\begin{equation}
\chi_n \in [0,1], \quad \rho_n \in [0,1] \tag{\ref{eq:delay_min}b}
\end{equation}

\begin{proposition}
The delay minimization problem is non-convex due to coupled variables in both the objective function and energy constraint.
\end{proposition}

To address this challenge, we introduce an auxiliary variable transformation.

\begin{lemma}[Problem Transformation]
By introducing $\delta_n^{\mathrm{tx}} = \frac{D_n[t]}{B \log_2(1 + \rho_n P_n |h_n|^2)}$, we can derive:
\begin{equation}
\rho_n = \frac{2^{\frac{D_n[t]}{\delta_n^{\mathrm{tx}} B}} - 1}{P_n |h_n|^2}.
\label{eq:rho_transform}
\end{equation}
\end{lemma}

The transformed problem becomes:

\begin{equation}
\min_{\chi_n,\delta_n^{\mathrm{tx}}} \frac{\mu \zeta_n}{\chi_n G_n} + \delta_n^{\mathrm{tx}}
\label{eq:transformed_problem}
\end{equation}

\begin{align*}
\text{s.t.} & \quad \kappa \mu \zeta_n (\chi_n G_n)^2 + \delta_n^{\mathrm{tx}} \frac{2^{\frac{D_n[t]}{\delta_n^{\mathrm{tx}} B}} - 1}{|h_n|^2} \leq e_n^{\max} \tag{\ref{eq:transformed_problem}a} \\
& \quad \frac{2^{\frac{D_n[t]}{\delta_n^{\mathrm{tx}} B}} - 1}{P_n |h_n|^2} \leq 1 \tag{\ref{eq:transformed_problem}b} \\
& \quad \chi_n \in [0,1] \tag{\ref{eq:transformed_problem}c}
\end{align*}

\begin{theorem}[Feasibility Condition]
The resource allocation problem is infeasible if $\ln(2)D_n[t] \geq e_n^{\max} B |h_n|^2$.
\end{theorem}

To derive the optimal solution, we apply the Lagrangian dual decomposition. We introduce four Lagrangian multipliers: $\lambda_1$ for the energy consumption constraint, $\lambda_2$ for the power allocation constraint, and $\lambda_3$, $\lambda_4$ for the computational resource allocation bounds. The Lagrangian function is:

\begin{small}
\begin{equation}
\begin{aligned}
\mathcal{L}(\chi_n, \delta_n^{\mathrm{tx}}, & \vec{\lambda})
= \frac{\mu \zeta_n}{\chi_n G_n} + \delta_n^{\mathrm{tx}} \\
&\mathllap{+} \lambda_1 \left(\kappa \mu \zeta_n (\chi_n G_n)^2 + \delta_n^{\mathrm{tx}} \frac{2^{\frac{D_n[t]}{\delta_n^{\mathrm{tx}} B}} - 1}{|h_n|^2} - e_n^{\max}\right) \\
&\mathllap{+} \lambda_2 \left(\frac{2^{\frac{D_n[t]}{\delta_n^{\mathrm{tx}} B}} - 1}{P_n |h_n|^2} - 1\right) - \lambda_3 \chi_n + \lambda_4 (\chi_n - 1)
\end{aligned}.
\end{equation}
\end{small}

According to the KKT conditions, we have:
\begin{small} 
\begin{equation}
\left\{
\begin{aligned}
\frac{\partial \mathcal{L}}{\partial \chi_n} &= -\frac{\mu \zeta_n}{\chi_n^2 G_n} + 2\lambda_1 \kappa \mu \zeta_n G_n^2 \chi_n - \lambda_3 + \lambda_4 = 0 \\
\frac{\partial \mathcal{L}}{\partial \delta_n^{\mathrm{tx}}} &= 1 + \lambda_1 \left[\frac{2^{\frac{D_n[t]}{\delta_n^{\mathrm{tx}} B}} - 1}{|h_n|^2} - \frac{D_n[t] \ln(2)}{B |h_n|^2 \delta_n^{\mathrm{tx}}} \cdot 2^{\frac{D_n[t]}{\delta_n^{\mathrm{tx}} B}}\right] \\
&\quad - \lambda_2 \frac{D_n[t] \ln(2)}{B P_n |h_n|^2 \delta_n^{\mathrm{tx}}} \cdot 2^{\frac{D_n[t]}{\delta_n^{\mathrm{tx}} B}} = 0
\end{aligned}
\right..
\end{equation}
\end{small}

Additionally, the complementary slackness conditions are:
\begin{small}
\begin{equation}
\left\{
\begin{aligned}
&\lambda_1 \left(\kappa \mu \zeta_n (\chi_n G_n)^2 + \delta_n^{\mathrm{tx}} \frac{2^{\frac{D_n[t]}{\delta_n^{\mathrm{tx}} B}} - 1}{|h_n|^2} - e_n^{\max}\right) = 0 \\
&\lambda_2 \left(\frac{2^{\frac{D_n[t]}{\delta_n^{\mathrm{tx}} B}} - 1}{P_n |h_n|^2} - 1\right) = 0 \\
&\lambda_3 \chi_n = 0, \quad \lambda_4 (\chi_n - 1) = 0 \\
&\lambda_i \geq 0, \quad \forall i \in \{1,2,3,4\}
\end{aligned}
\right. .
\end{equation}
\end{small}

From the KKT condition $\frac{\partial \mathcal{L}}{\partial \chi_n} = 0$, we have:
\begin{equation}
-\frac{\mu \zeta_n}{\chi_n^2 G_n} + 2\lambda_1 \kappa \mu \zeta_n G_n^2 \chi_n - \lambda_3 + \lambda_4 = 0.
\end{equation}

To derive the optimal resource allocation coefficient, we analyze different cases based on the complementary slackness conditions:
\begin{equation}
\lambda_3 \chi_n = 0,\ \lambda_4 (\chi_n - 1) = 0,\ \lambda_i \geq 0,\ \forall i \in \{1,2,3,4\}
\end{equation}

\textbf{Case 1:} When $\lambda_1 = 0$ (energy constraint not binding).

With $\lambda_1 = 0$, the KKT condition simplifies to:
\begin{equation}
-\frac{\mu \zeta_n}{\chi_n^2 G_n} - \lambda_3 + \lambda_4 = 0.
\end{equation}

If we assume $\chi_n < 1$, then $\lambda_4 = 0$, which gives:
\begin{equation}
-\frac{\mu \zeta_n}{\chi_n^2 G_n} - \lambda_3 = 0.
\end{equation}

This implies $\lambda_3 = -\frac{\mu \zeta_n}{\chi_n^2 G_n} < 0$, contradicting the non-negativity constraint on $\lambda_3$.

Therefore, we must have $\chi_n = 1$, with $\lambda_3 = 0$ and $\lambda_4 \geq 0$.

\textbf{Case 2:} When $\lambda_1 > 0$ (energy constraint binding).

We return to the original KKT condition:
\begin{equation}
-\frac{\mu \zeta_n}{\chi_n^2 G_n} + 2\lambda_1 \kappa \mu \zeta_n G_n^2 \chi_n - \lambda_3 + \lambda_4 = 0.
\end{equation}

We consider three subcases:

\textbf{Subcase 2.1:} $\chi_n = 0$.

This gives $-\infty + 0 - \lambda_3 + \lambda_4 = 0$, which is mathematically infeasible. Thus, $\chi_n \neq 0$.

\textbf{Subcase 2.2:} $0 < \chi_n < 1$.

By complementary slackness, $\lambda_3 = 0$ and $\lambda_4 = 0$, yielding:
\begin{equation}
-\frac{\mu \zeta_n}{\chi_n^2 G_n} + 2\lambda_1 \kappa \mu \zeta_n G_n^2 \chi_n = 0.
\end{equation}

Rearranging the above equation, we have:
\begin{equation}
\frac{\mu \zeta_n}{\chi_n^2 G_n} = 2\lambda_1 \kappa \mu \zeta_n G_n^2 \chi_n.
\end{equation}

Further simplification yields:
\begin{equation}
\frac{1}{\chi_n^3} = 2\lambda_1 \kappa G_n^3.
\end{equation}

Solving for $\chi_n$:
\begin{equation}
\chi_n = \left(\frac{1}{2\lambda_1 \kappa G_n^3}\right)^{\frac{1}{3}} = \left(\frac{\mu \zeta_n}{2\lambda_1 \kappa \mu \zeta_n G_n^3}\right)^{\frac{1}{3}}.
\end{equation}

Note that $\mu$ and $\zeta_n$ appear in both numerator and denominator, allowing them to cancel out.

\textbf{Subcase 2.3:} $\chi_n = 1$.

With $\lambda_3 = 0$ and $\lambda_4 \geq 0$, we need to verify this solution satisfies the original KKT condition.

Combining all cases and considering the constraint $\chi_n \in [0,1]$, we can obtain the optimal solution through the following theorem.

\begin{theorem}[Optimal Resource Allocation]
\label{thm:optimal_resource}
The optimal resource allocation strategy is:

If $\lambda_1 = 0$ (energy constraint not binding):
   \begin{equation}
   \chi_n^* = 1, \quad \rho_n^* = 1
   \label{eq:opt_no_constraint}
   \end{equation}

If $\lambda_1 > 0$ (energy constraint binding):
   \begin{equation}
   \chi_n^* = \min\left\{\left(\frac{\mu \zeta_n}{2\lambda_1 \kappa \mu \zeta_n G_n^3}\right)^{\frac{1}{3}}, 1\right\} ,
   \label{eq:opt_chi}
   \end{equation}
   and 
   \begin{equation}
   \rho_n^* = \min\left\{\frac{2^{\frac{D_n[t]}{\delta_n^{\mathrm{tx}*} B}} - 1}{P_n |h_n|^2}, 1\right\},
   \label{eq:opt_rho}
   \end{equation}
where $\delta_n^{\mathrm{tx}*}$ satisfies:
   \begin{equation}
   \kappa \mu \zeta_n (\chi_n^* G_n)^2 + \delta_n^{\mathrm{tx}*} \frac{2^{\frac{D_n[t]}{\delta_n^{\mathrm{tx}*} B}} - 1}{|h_n|^2} = e_n^{\max}.
   \label{eq:opt_delta}
   \end{equation}
\end{theorem}

Note that in Eq. (\ref{eq:opt_chi}), the parameters $\mu$ and $\zeta_n$ appear in both the numerator and denominator of the fraction, allowing them to cancel out. This mathematical simplification yields the final expression without these parameters. The optimal value of $\delta_n^{\mathrm{tx}*}$ in Eq. (\ref{eq:opt_delta}) typically requires numerical methods to solve, such as one-dimensional search or numerical root-finding techniques, as the equation $\delta_n^{\mathrm{tx}} \cdot 2^{(D_n[t]/\delta_n^{\mathrm{tx}}B)}$ does not admit a simple closed-form solution.

The optimal solution provides the resource allocation for each PF, ensuring minimum delay while adhering to energy constraints. This analytical approach offers significant advantages over exhaustive search methods, as it provides insights into the resource allocation trade-offs and substantially reduces computational complexity.

\textcolor{blue}{To address scalability concerns with large neural networks, we provide approximations and closed-form solutions for specific scenarios.}

\textcolor{blue}{\begin{lemma}[Large Model Approximation]\label{lemma:large_model}
When $\frac{D_n[t]}{\delta_n^{\mathrm{tx}} B} \geq 3$, the energy constraint can be approximated as:
\begin{equation}
\kappa \mu \zeta_n (\chi_n^* G_n)^2 + \delta_n^{\mathrm{tx}} \frac{e^{\frac{D_n[t] \ln(2)}{\delta_n^{\mathrm{tx}} B}}}{|h_n|^2} \approx e_n^{\max},
\end{equation}
which yields the closed-form approximation:
\begin{equation}
\delta_n^{\mathrm{tx}*} \approx \frac{D_n[t] \ln(2)}{B \ln\left(\frac{(e_n^{\max} - \kappa \mu \zeta_n (\chi_n^* G_n)^2)|h_n|^2}{D_n[t] \ln(2)}\right)}.
\end{equation}
\end{lemma}}

\textcolor{blue}{\begin{theorem}[High SNR Solution]\label{thm:high_snr}
When $P_n |h_n|^2 \geq 10$ and the energy constraint is binding, the optimal transmission time has the closed-form solution:
\begin{equation}
\delta_n^{\mathrm{tx}*} = \frac{D_n[t]}{B \log_2\left(1 + \frac{e_n^{\max} |h_n|^2}{\kappa \mu \zeta_n (\chi_n^* G_n)^2}\right)}.
\end{equation}
\end{theorem}}
\begin{proof}
\textcolor{blue}{The detailed proof is provided in Appendix A.}
\end{proof}

\textcolor{blue}{These approximations reduce computational complexity from $O(I \log(\epsilon^{-1}))$ for iterative numerical methods to $O(I)$ for direct evaluation, representing significant computational savings while maintaining solution accuracy within 5\% of the optimal numerical solution.}

\subsection{Multi-Agent Reinforcement Learning-Based Solution for Vehicle Selection}

With the optimal resource allocation obtained from Section \ref{Method}-A, we now focus on selecting appropriate vehicles to minimize the weighted sum of AoI and FLMD as formulated in (\ref{eq:AoI_FLMD}). To address the high-dimensional state-action space challenge, we adopt a multi-agent reinforcement learning approach where each agent $k \in \mathcal{K}$ controls one subchannel.

\subsubsection{$M$-Order Markov Decision Process in WFL-empowered VP}

To capture the temporal dynamics of platoon motion, we formulate the problem as an $M$-th order MDP. Specifically, we partition each communication round into $M$ sub-periods, with state $S[t] = \{s_1[t], s_2[t], \ldots, s_M[t]\}$.

The state vector for each sub-period $m$ contains three key elements for each device $n$:
\begin{equation}
s_{m}[t] = \{(\Theta_n[t], |h_n|^2, \Delta_n[t]) | n \in \mathcal{N}\},
\end{equation}
where $\Theta_n[t]$ is the FLMD as defined in Theorem \ref{thm:flmd_bound}, $|h_n|^2$ is the current channel gain, and $\Delta_n[t]$ is the cumulative Age of Information for device $n$.

For the FLMD-based selection, we define a fixed-dimension binary mask vector $\mathbf{m}_t \in \{0,1\}^N$ where:
\begin{equation}
\mathbf{m}_t[n] = 
\begin{cases}
1, & \text{if } \Theta_n[t] \leq \Lambda_{\Theta} \\
0, & \text{otherwise}
\end{cases}.
\end{equation}

This mask is incorporated in the state representation and used to mask the action probabilities during policy evaluation:
\begin{equation}
\pi_{\theta}(a|s,\mathbf{m}) = \text{softmax}(f_{\theta}(s) \odot \mathbf{m}),
\end{equation}
where $\odot$ represents element-wise multiplication, ensuring that ineligible devices (with $\mathbf{m}[n]=0$) have zero probability of being selected.

Based on Theorem \ref{thm:convergence}, which shows that the convergence rate is affected by the term $(1-\frac{\mu}{L})^t$, we propose an adaptive probability mask that balances exploration and exploitation:

\begin{equation}
\tilde{\mathbf{m}}_t[n] = 
\begin{cases}
1, & \text{if}~\Theta_n[t] \leq \Lambda_{\Theta} \\
e^{-\beta\Theta_n[t](1-\frac{\mu}{L})^{-t}}, & \text{otherwise}
\end{cases}.
\label{eq:adaptive_mask}
\end{equation}

where $\beta$ is a temperature parameter. This formulation allows devices with slightly higher FLMD to still have a non-zero probability of being selected, especially in later rounds where the impact of model drift diminishes according to our theoretical analysis.

\subsubsection{Temporal Sequence Feature Extraction Network}

To effectively process the temporal dynamics captured by our $M$-th order MDP, we introduce the Temporal Sequence Feature Extraction Network (TSFEN). The architecture comprises an MHSA layer for capturing inter-device relationships, followed by an LSTM layer for temporal feature extraction, and a fully connected layer for final output generation:

\begin{equation}
\text{TSFEN}(S[t]) = \text{FC}(\text{LSTM}(\text{MHSA}(S[t]))).
\end{equation}

The MHSA mechanism computes multiple attention heads that operate on the input state $S[t]$, with each head focusing on different aspects of the device relationships. The LSTM layer then processes these attention-weighted features to capture temporal dependencies across the $M$ sub-periods. The adaptive mask $\tilde{\mathbf{m}}_t$ derived from our theoretical analysis in Theorem \ref{thm:convergence} is integrated into the final action selection process:

\begin{equation}
\pi_{\theta}(a|S[t]) = \text{softmax}(\text{FC}(h_M) \odot \tilde{\mathbf{m}}_t).
\end{equation}
where $h_M$ is the final hidden state output from the LSTM layer at the $M$-th step. This architecture effectively leverages both spatial relationships among devices and temporal patterns across communication rounds, enabling informed vehicle selection decisions while maintaining the theoretical guarantees established in Section \ref{ProFm}.

\subsubsection{Markov Decision Process}

\begin{algorithm}[t]
\caption{MAPPO for Vehicle Selection}
\label{alg:MAPPO}
\begin{algorithmic}[1]
\State \textbf{Input:} Total episodes $E$; total communication round $T$; total agent $K$; total vehicles $N$; learning rate $\beta$; discount factor $\gamma$; advantage decay $\lambda$;
\State \textbf{Output:} The actor network parameters $\theta$, the critic network parameters $\psi$
\For{each episode $e=1, \ldots, E$}
    \State Initialize the VP environment
    \For{each communication round $t=1, \ldots, T$}
        \For{each vehicle $n=1, \ldots, N$}
            \State Obtained from Lagrangian Dual Decomposition (Theorem~\ref{thm:optimal_resource}), denoted as $\chi_n^*$, $\rho_n^*$
        \EndFor
        \For{each agent $k=1, \ldots, K$}

            \State Observe state $S_k[t]$ from the environment
            \State Select action $A_k[t]$ according to policy $\pi_{\theta_k}(S_k[t])$
            \State Execute action $A_k[t]$, observe reward $R_k[t]$ and new state $S_k[t+1]$
            \State Calculate TD residual by (\ref{TD})
            \State Calculate the value loss by (\ref{critic loss}) and update Critic network parameters by (\ref{critic update})
            \State Calculate the advantage estimate by (\ref{GAE})
            \State Calculate the probability ratio by (\ref{probability ratio}) and the clipped ratio by (\ref{Clipped Probability Ratio})
            \State Calculate policy gradient by (\ref{policy gradient}) and Update actor network by (\ref{actor update})
        \EndFor
    \EndFor
\EndFor
\end{algorithmic}
\end{algorithm}

\begin{itemize}
    \item \textbf{State Space:} 
    In the $t$-th communication round, the $k$-th agent possesses $M$ states, each representing the real-time state for a sub-period within that round. The state vector for each sub-period $m$ contains three key elements for each device:
    \begin{equation}
    s_{k,m}[t] = \{(\Theta_n[t], |h_n|^2, \Delta_n[t]) | n \in \mathcal{N}\}.
    \end{equation}
    
    Consequently, the state of the $k$-th agent during the $t$-th communication round is:
    \begin{equation}
    S_k[t] = \{s_{k,1}[t], \dots, s_{k,m}[t], \dots, s_{k,M}[t]\}.
    \end{equation}
    
    \item \textbf{Action Space:} 
    Since each agent represents a corresponding sub-channel to facilitate vehicle selection, the action space is defined as the indicators of the vehicles selected at the start of each communication round:
    \begin{equation}
    A_k[t] = \{\phi_{1,k}[t], \phi_{2,k}[t], \ldots, \phi_{N,k}[t]\},
    \end{equation}
    where $\phi_{n,k}[t]$ is an indicator variable that denotes whether the $n$-th vehicle is selected by the $k$-th sub-channel in the $t$-th communication round.
    
    For the FLMD-based selection, we define a fixed-dimension binary mask vector $\mathbf{m}_t \in \{0,1\}^N$ where:
    \begin{equation}
    \mathbf{m}_t[n] = 
    \begin{cases}
    1, & \text{if } \Theta_n[t] \leq \Lambda_{\Theta} \\
    0, & \text{otherwise}
    \end{cases}.
    \end{equation}
    
    This mask is incorporated in the state representation and used to mask the action probabilities during policy evaluation:
    \begin{equation}
    \pi_{\theta}(a|s,\mathbf{m}) = \text{softmax}(f_{\theta}(s) \odot \mathbf{m}),
    \end{equation}
    where $\odot$ represents element-wise multiplication, ensuring that ineligible devices (with $\mathbf{m}[n]=0$) have zero probability of being selected.

    \item \textbf{Rewards Function:} 
    Based on our AoI optimization problem in (\ref{eq:AoI_FLMD}), we define a reward function that jointly considers AoI and FLMD:
    \begin{equation}
    R_k[t] = -\frac{\alpha \sum_{n=1}^N \Delta_n[t] + \beta \sum_{n=1}^N \Theta_n[t]^2}{M \cdot K}.
    \end{equation}
    where $\alpha$ and $\beta$ are positive weighting coefficients that balance the importance between AoI optimization and FLMD reduction. These parameters allow for flexible adjustment of the learning focus based on system requirements: increasing $\alpha$ prioritizes information freshness, while higher $\beta$ values emphasize model consistency and security.
    
    Drawing from Theorem \ref{thm:convergence}, which shows that convergence is affected by $\mathbb{E}[\|e[t]\|^2]$, and Lemma \ref{lemma:deviation}, which relates this to FLMD, we use $\Theta_n[t]^2$ rather than $\Theta_n[t]$ to directly optimize the convergence bound.
\end{itemize}

By extending the state space to include FLMD-related information and adjusting the reward function to penalize large $\Theta_n[t]$, our MAPPO framework can learn to effectively balance AoI minimization with model security. Combining this FLMD-aware device selection with the prior grid-search resource allocation ensures that each PF's training remains both efficient and robust, providing a comprehensive optimization solution in the WFL-empowered VP scenario.

\textcolor{blue}{The MAPPO algorithm for vehicle selection is shown in Algorithm \ref{alg:MAPPO}. The computational complexity is dominated by the number of agents, neural network architecture, and training iterations. Each agent's state observation and action selection requires $O(n \cdot m^2)$ complexity for an $n$-layer network with $m$ neurons per layer. The overall complexity for $E$ episodes with $T$ communication rounds and $N$ agents is $O(E \cdot T \cdot N \cdot (n \cdot m^2 + N_{batch} \cdot n \cdot m^2))$.}

\subsection{Training Algorithm}

At the beginning of training MAPPO, each agent observes the state $S_k[t]$ from the environment, and an action  $A_k[t]$ is then selected according to the policy $\pi_{\theta_k}$. $A_k[t]$ is executed, and the agent observes the reward $R_k[t]$ and the new state $S_k[t+1]$.
 Specifically, the implementation of this algorithm involves designating each sub-channel as an agent and training two networks: the critic network and the actor network. The critic network learns a mapping function $V_\psi$ from the state space $S$ to a real-valued reward $R$, while the actor network learns to sample from the action distribution $a$ based on observations $O(t)$ through the mapping function $\pi_\phi$.

Following the observation of the state \(S_k[t]\), each agent selects an action \(A_k[t]\) based on the policy \(\pi_{\theta_k}\), which is parameterized by \(\theta_k\). Upon executing the action, the agent observes a reward \(R_k[t]\) and transitions to a new state \(S_k[t+1]\). The temporal difference (TD) residual is then computed as:
\begin{equation}
\epsilon_k[t]=R_k[t]+\gamma V_{\psi_k}\left(S_k[t+1]\right)-V_{\psi_k}\left(S_k[t]\right),
\label{TD}
\end{equation}
where \(\gamma\) is the discount factor, and \(V_{\psi_k}\) represents the value function estimated by the critic network, parameterized by \(\psi_k\).

In RL, the objective of an agent is to learn and optimize its policy through interactions with the environment, aiming to maximize the cumulative rewards. Specifically, within the MAPPO, the reward function is used to evaluate the relative benefit of taking a particular action in the current state compared to alternative actions, indicating the extent to which an action is expected to yield higher or lower returns compared to others. During the process of policy optimization, MAPPO incorporates Generalized Advantage Estimation (GAE) to achieve a more accurate estimation of the advantage function:
\begin{equation}
\hat{A}_k[t]=\epsilon_k[t]+\gamma \lambda \hat{A}_k[t+1],
\label{GAE}
\end{equation}
where \(\epsilon_k[t]\) is the TD residual, \(\lambda\) is the decay parameter for the advantage update, and \(\gamma\) is the discount factor. This recursive formulation of GAE helps in smoothing out the variance in advantage estimates, making it particularly effective for environments with high variability in rewards.

To update the critic network, the value loss \(L_{\psi_k}\) is calculated, which typically uses the Mean Squared Error (MSE) between the estimated values and the true values of states. The formula for the value loss \(L_{\psi_k}\) is given by:
\begin{equation}
L_{\psi_k}=\frac{1}{2} \mathbb{E}\left[\left(\epsilon_k[t]\right)^2\right],
\label{critic loss}
\end{equation}
where \(\epsilon_k[t]\) is the TD residual for agent \(k\) at time \(t\). Following the calculation of the value loss, the parameters \(\psi_k\) of the critic network are adjusted accordingly:
\begin{equation}
\psi_k \leftarrow \psi_k-\beta \nabla_{\psi_k} L_{\psi_k},
\label{critic update}
\end{equation} 
where \(\beta\) is the learning rate. Subsequently, the policy update involves calculating the probability ratio \(r_k[t]\), reflecting the likelihood of selecting action \(A_k[t]\) under the current policy relative to the baseline policy:
\begin{equation}
r_k[t]=\frac{\pi_{\theta_k}\left(A_k[t] \mid S_k[t]\right)}{\pi_{\theta_{\text {old }}}\left(A_k[t] \mid S_k[t]\right)}.
\label{probability ratio}
\end{equation}

This ratio is clipped to ensure stable policy updates:
\begin{equation}
L_k[t]=\operatorname{clip}\left(r_k[t], 1-\epsilon, 1+\epsilon\right),
\label{Clipped Probability Ratio}
\end{equation}
where \(\epsilon\) is a small positive number to restrict the extent of modification to the policy ratio. The policy gradient is then computed using the clipped ratio, which helps to minimize the risk of large policy updates that can lead to performance degradation:
\begin{equation}
\nabla_{\theta_k} J\left(\theta_k\right)=\mathbb{E}\left[\min \left(r_k[t] \cdot \hat{A}_k[t], L_k[t] \cdot \hat{A}_k[t]\right)\right],
\label{policy gradient}
\end{equation}
and the actor network parameters \(\theta_k\) are updated:
\begin{equation}
\theta_k \leftarrow \theta_k+\beta \nabla_{\theta_k} J\left(\theta_k\right).
\label{actor update}
\end{equation}

\section{Performance Evaluation} \label{PerfEvl}

\subsection{Parameter Settings}

\begin{table}[t]
\centering
\caption{Parameter settings for simulations.}
\label{tab:simulation_params}
\begin{tabular}{>{\arraybackslash}p{5cm} >{\arraybackslash}p{3cm}}
  \toprule
  \textbf{Parameter} & \textbf{Value} \\ 
  \midrule
  Number of PFs ($N$),  & 20 \\ 
  Path loss exponent ($\alpha$) & 3.76 \\
  CPU cycles for each bit of tasks ($\mu$) & $10^7$\\
  Model size ($D$) & 1 Mbit \textcolor{blue}{(transmitted gradients)}\\
  Available CPU cycles ($G_n$) & $0.5$ ${\rm GHz}$ \\
  Maximum transmit power ($P_n$) & $15$ ${\rm dBm}$ \\
  Maximum energy consumption ($e_n^{\max }$) & $0.1$ ${\rm Joule}$ \\ 
  Power consumption coefficient ($\kappa$) & $10^{-28}$ \\
  Maximum acceleration ($a_{max}$) & $0.73$ ${\rm m/s^2}$ \\
  Maximum deceleration ($b_{\max}$ ) & $1.67$ ${\rm m/s^2}$ \\
  Minimum inter-vehicle space ($d_{\min}$ ) &  $2.0$ ${\rm m}$\\
  Minimum reaction time ($t_{\min}$) & $1.5$ ${\rm s}$\\
  Desired velocity ($v_{\text{des}}$) & $30 $ ${\rm m/s^2}$ \\
  Update interval ($\tau$) & $1$ ${\rm s}$\\
  Noise variance ($\sigma^2$) & $-174$ ${\rm dBm}$\\
  Batch size ($N_{batch}$) & 32 \\
  Learning rate in AI4MARS ($\xi$) & $10^{-4}$ \\
  Learning rate in MAPPO/MADDQN ($\beta$) &  $10^{-4}$/ $10^{-4}$ \\
  Discount factor in MAPPO/MADDQN ($\gamma$) & $0.98$/ $0.95$ \\
  Replay buffer ($\mathcal{B}$) & $100000$ \\
  Optimizer in AI4MARS &  AdamW\\
  Optimizer in MAPPO/MADDQN & Adam/ Adam \\

  \bottomrule
\end{tabular}
\end{table}

To verify our proposed approach, we investigate the performance of RACE, including Lagrangian Dual Decomposition-based resource management and MAPPO-based vehicle selection algorithm in the WFL-empowered VP system via simulations. We compare our approach with Multi-Agent Double Deep Q-Network (MADDQN) as a baseline. Furthermore, we evaluate the impact of different neural network architectures by implementing both the proposed TSFEN with MHSA and LSTM components, and a standard Multi-Layer Perceptron (MLP) structure for each method. The system in the simulations consists of a PL and $N$ PFs. Each PF's initial speed and inter-vehicle distance are randomly generated within ranges of $[15, 20]$ m/s and $[10, 15]$ m, respectively. The length of each vehicle is set to $5$ m. The parameter settings are shown in Table \ref{tab:simulation_params}. \textcolor{blue}{We have expanded our evaluation to three comprehensive scenarios: $N=10$, $K=2$ (small platoons for highway trucking convoys), $N=20$, $K=4$ (medium platoons for standard deployments), and $N=30$, $K=6$ (large platoons for dense urban environments). We have also implemented a Convex-Greedy baseline that combines CVX/MOSEK convex optimization with AoI-based greedy selection, representing state-of-the-art approaches commonly used in federated learning literature.}

The experiments utilize the AI4MARS dataset, which contains terrain annotations for Mars rover imagery \cite{Mar2021CVPR}. This dataset consists of approximately 326K semantic segmentation labels on 35K images captured by the Curiosity (MSL), Opportunity, and Spirit (MER) Mars rovers. The dataset includes 17K NAVCAM images and 9K Mastcam images from Curiosity, 6K NAVCAM images from Opportunity, and 3K NAVCAM images from Spirit. The images are classified into four terrain types critical for Mars rover traversability: Soil (36.32\%), Bedrock (46.09\%), Sand (15.61\%), and Big Rock (1.97\%). Each image in the training set was labeled by approximately 10 different annotators through a crowdsourcing approach to ensure labeling quality and consensus. The annotation process employed a pixel-wise merging strategy where a terrain class was accepted only if at least three different annotators agreed on the label with a minimum of 65\% consensus. For evaluation, we use a separate test set containing approximately 1.5K expert-labeled annotations created by rover planners and scientists from NASA's MSL and MER missions.

Due to the substantial computational demands of training on the entire dataset and to enable faster experimentation iterations, we employed a stratified sampling approach to extract a representative subset. Specifically, we randomly selected 30\% of the images from each terrain type, maintaining the original class distribution to ensure proper representation of all classes, including the rare Big Rock class. This resulted in a training set of approximately 10.5K images with 98K labels, which significantly reduced training time while preserving the dataset heterogeneity and class balance characteristics.

\begin{figure*}[t]
\centering
\begin{subfigure}[b]{0.3\textwidth}
    \includegraphics[width=\textwidth]{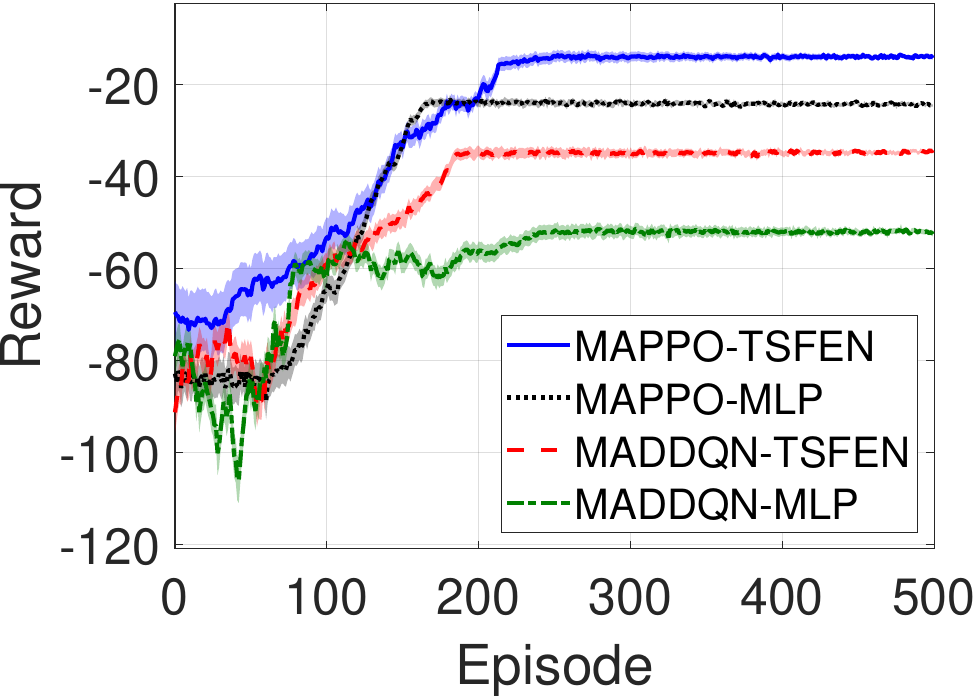}
    \caption{$K=2$}
\end{subfigure}
\hfill
\begin{subfigure}[b]{0.3\textwidth}
    \includegraphics[width=\textwidth]{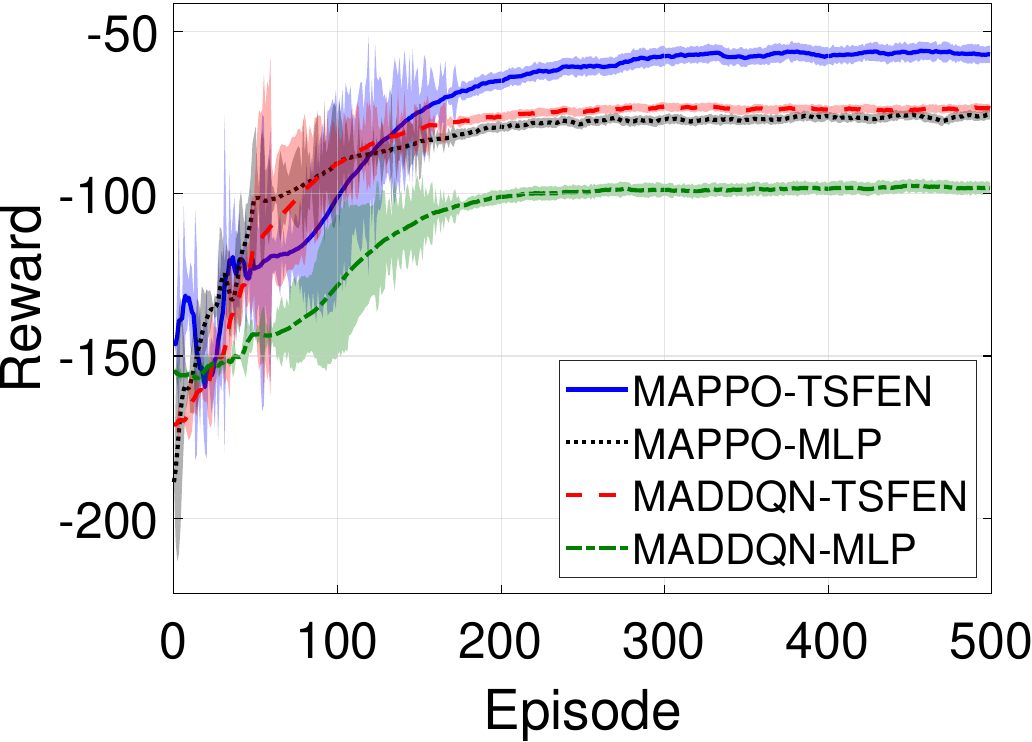}
    \caption{$K=4$}
\end{subfigure}
\hfill
\begin{subfigure}[b]{0.3\textwidth}
    \includegraphics[width=\textwidth]{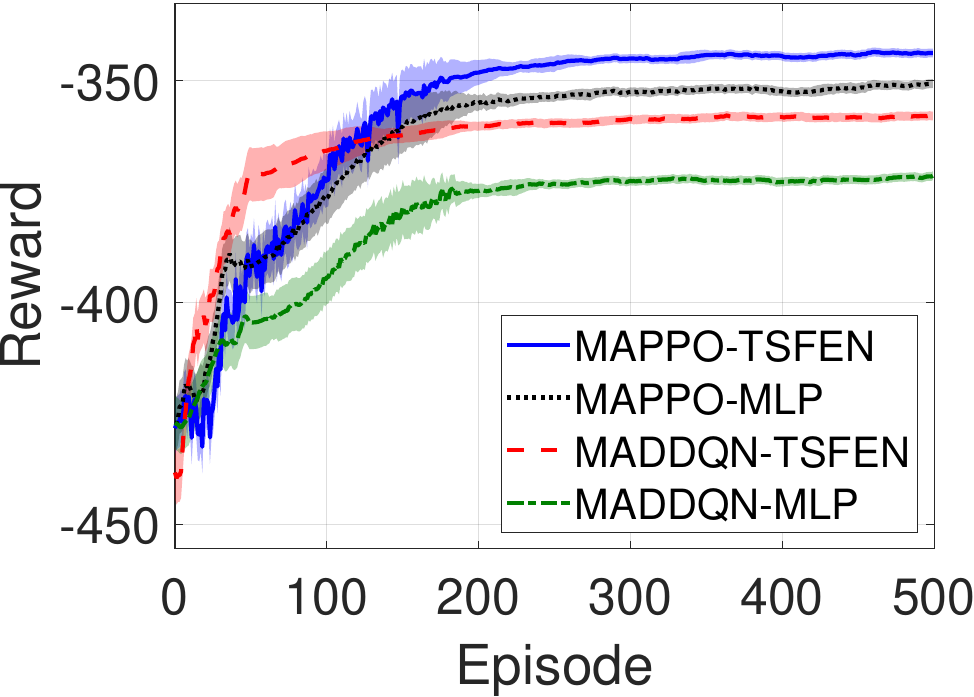}
    \caption{$K=6$}
\end{subfigure}
\caption{\textcolor{blue}{Reward convergence comparison of different reinforcement learning methods and network architectures with varying numbers of subchannels}.}
\label{Reward}
\end{figure*}


We implement a DeepLabv3+ architecture with a ResNet-101 backbone pretrained on ImageNet. The model employs atrous spatial pyramid pooling (ASPP) with dilation rates of 6, 12, and 18 to capture multi-scale context information. Images are resized to 513$\times$513 pixels for both training and inference. To address the significant class imbalance, particularly the under-represented Big Rock class (1.97\%), we employ a weighted loss function strategy. The weights are assigned inversely proportional to the class frequency in the training set, specifically using a 1-composition approach, where composition refers to the percentage of pixels belonging to each class.

\textcolor{blue}{The model size parameter represents transmitted gradient updates rather than full model parameters. Our DeepLabv3+ implementation uses a frozen ResNet101 backbone with only task-specific layers (ASPP module and decoder, $\sim$2.1M parameters) participating in federated aggregation. Through 16-bit quantization and gradient compression, the transmission size is reduced to approximately 1 Mbit per communication round.}

During training, we use the AdamW optimizer with a learning rate of 1e-4 and a polynomial decay schedule with power 0.9 over 200 epochs. Data augmentation techniques include random horizontal and vertical flips and color jittering to enhance the model's generalization capability and robustness to lighting variations, which are common in Mars imagery. We utilize the overall mIoU as our primary evaluation metric during the training process, as it provides a balanced assessment of segmentation performance across all terrain classes, particularly for imbalanced datasets. To simulate heterogeneous data distribution across the vehicle platoon, we employ a Dirichlet distribution (Dir($\phi$)) to create Non-IID data partitions. Specifically, for parameter $\phi=0.5$, we generate a probability vector $\mathbf{p} \sim \text{Dir}(\phi)$ for $N$ vehicles, and then allocate data from each class to different vehicles according to these probabilities, thereby ensuring that the training data received by each vehicle exhibits.

\textcolor{blue}{Our RACE framework is designed for realistic vehicular edge computing platforms using ARM Cortex-A76 processors with 8 GB RAM. System-level analysis demonstrates feasibility with total memory usage of 220 MB (2.7\% of available capacity), per-round latency under 60 ms (below 100 ms real-time constraint), and energy consumption of 4.2 J per round (less than 2\% hourly battery impact).}

\textcolor{blue}{To ensure reproducibility, our complete implementation specifications including network architectures, hyperparameters, and hardware requirements are provided in the supplementary materials.}

\begin{figure*}[htbp]
\centering
\begin{subfigure}[b]{0.3\textwidth}
    \includegraphics[width=\textwidth]{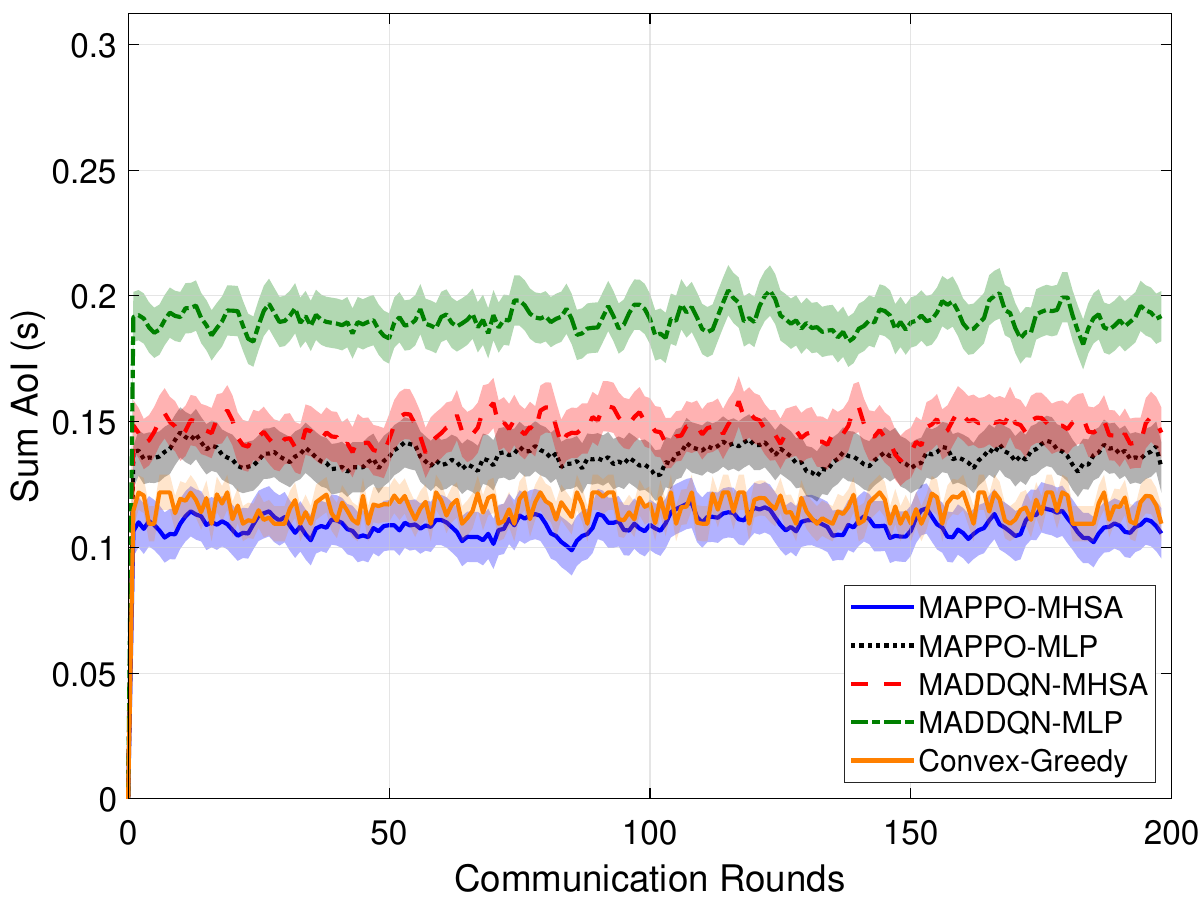}
    \caption{$K=2$}
\end{subfigure}
\hfill
\begin{subfigure}[b]{0.3\textwidth}
    \includegraphics[width=\textwidth]{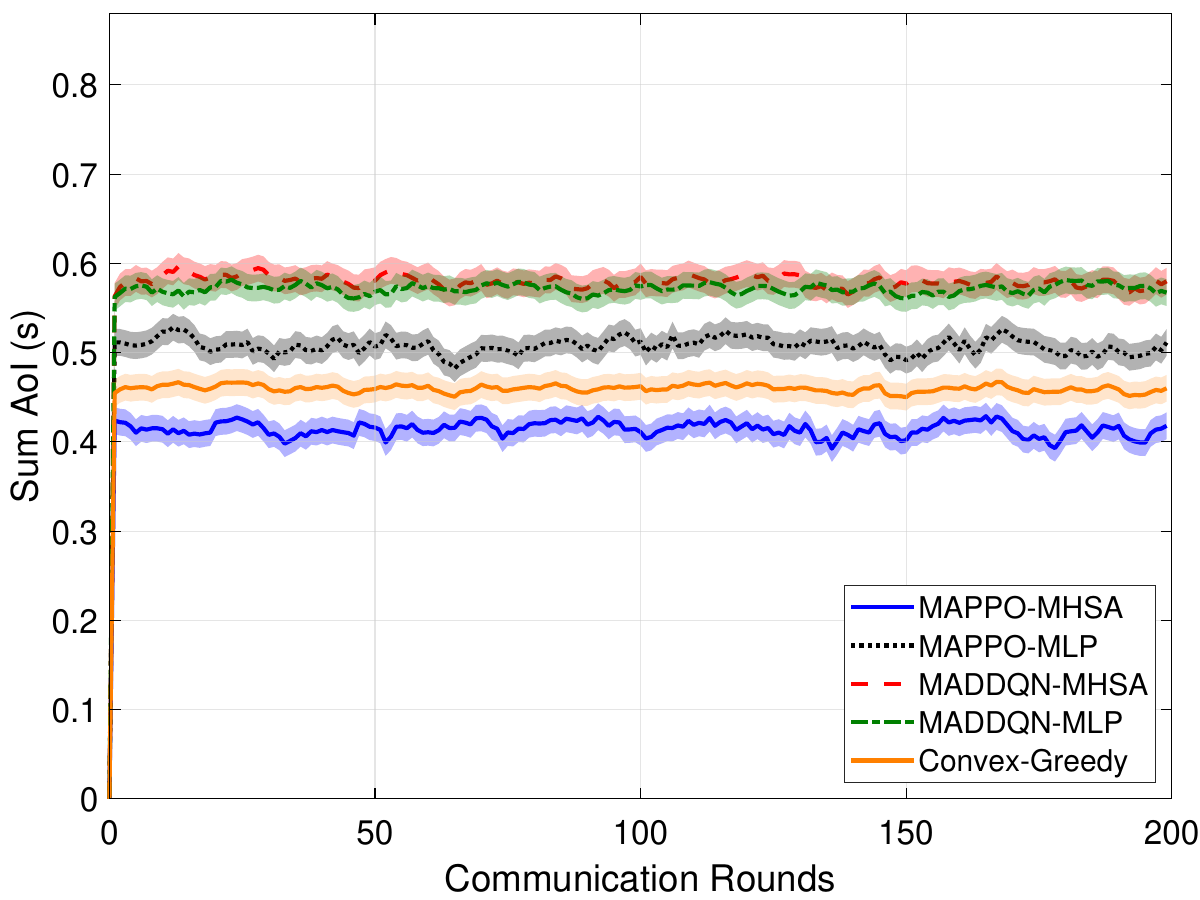}
    \caption{$K=4$}
\end{subfigure}
\hfill
\begin{subfigure}[b]{0.3\textwidth}
    \includegraphics[width=\textwidth]{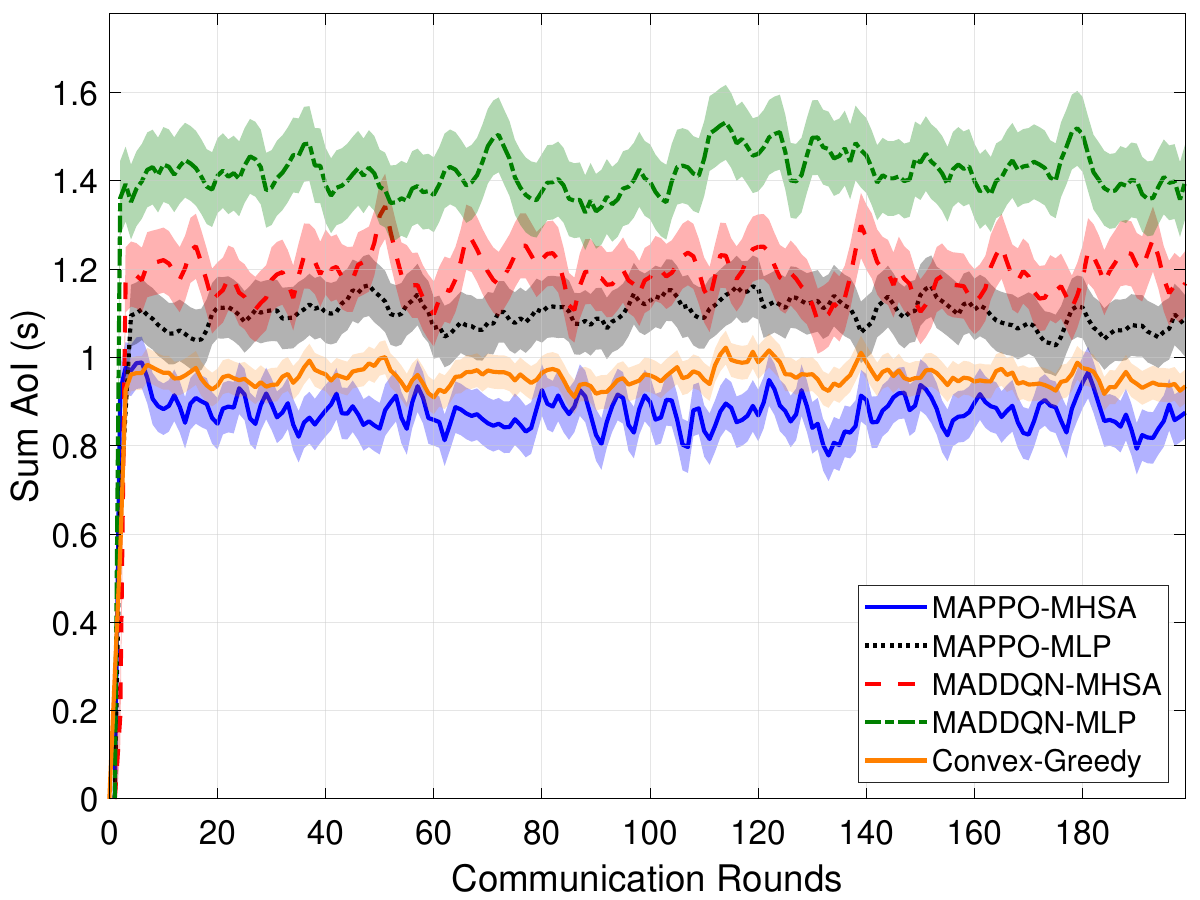}
    \caption{$K=6$}
\end{subfigure}
\caption{\textcolor{blue}{Comparison of AoI optimization performance across different methods and network architectures.}}
\label{AoI}
\end{figure*}

\begin{figure*}[htbp]
\centering
\begin{subfigure}[b]{0.3\textwidth}
    \includegraphics[width=\textwidth]{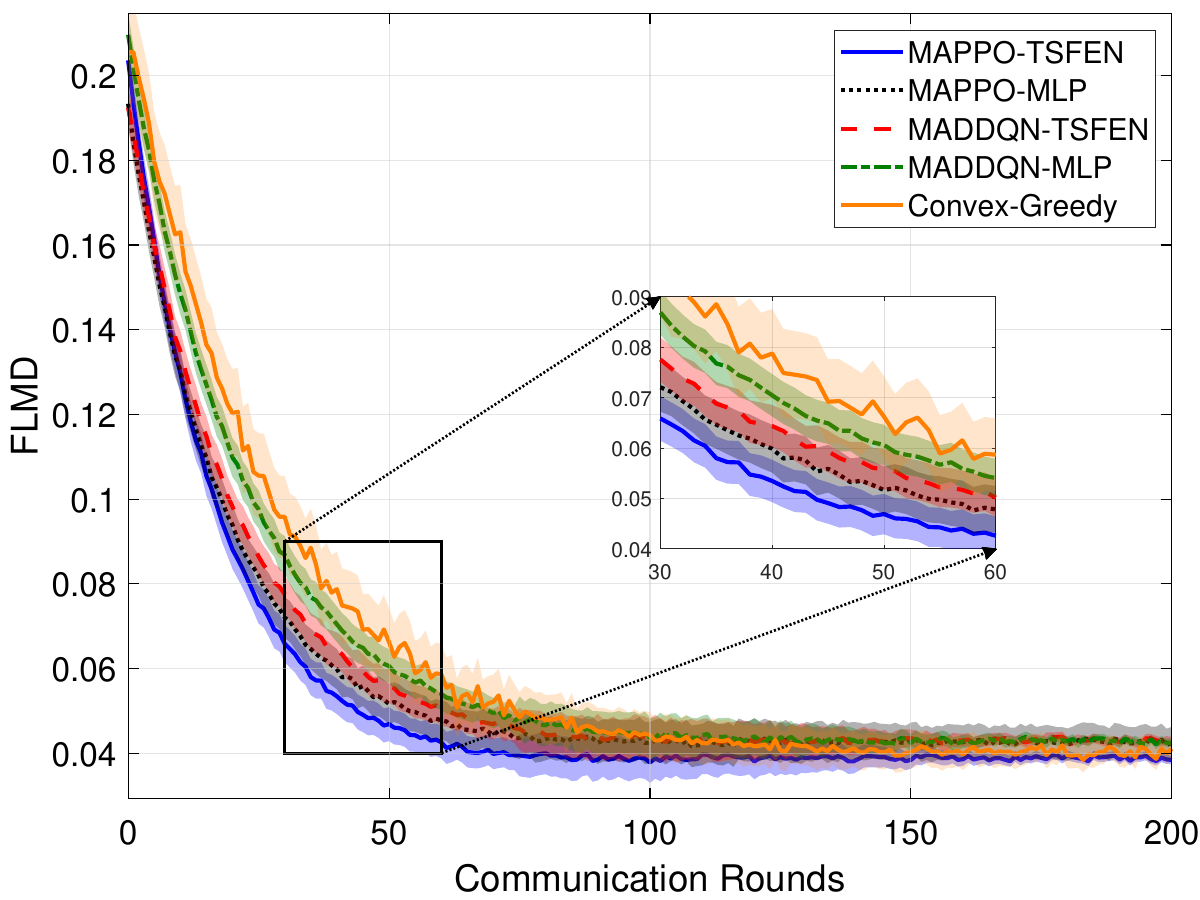}
    \caption{$K=2$}
\end{subfigure}
\hfill
\begin{subfigure}[b]{0.3\textwidth}
    \includegraphics[width=\textwidth]{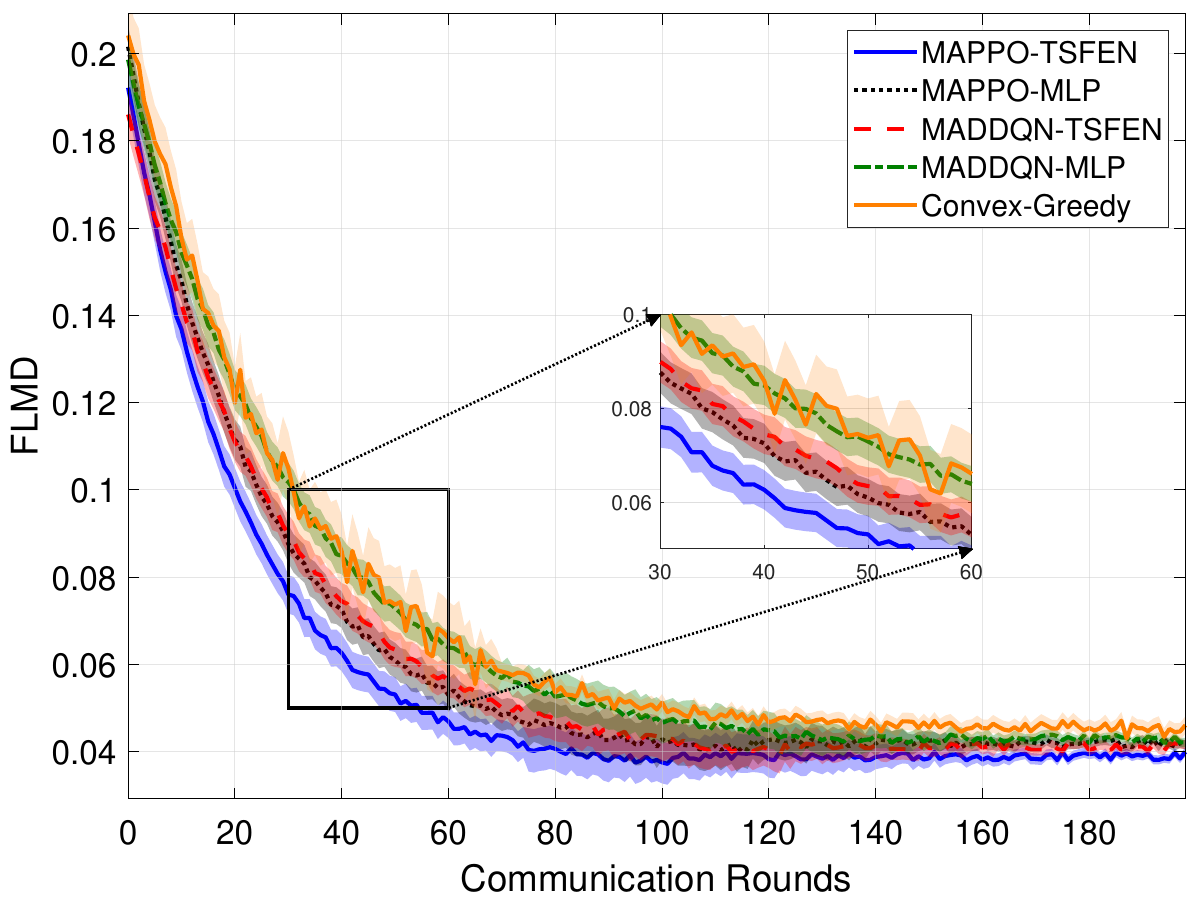}
    \caption{$K=4$}
\end{subfigure}
\hfill
\begin{subfigure}[b]{0.3\textwidth}
    \includegraphics[width=\textwidth]{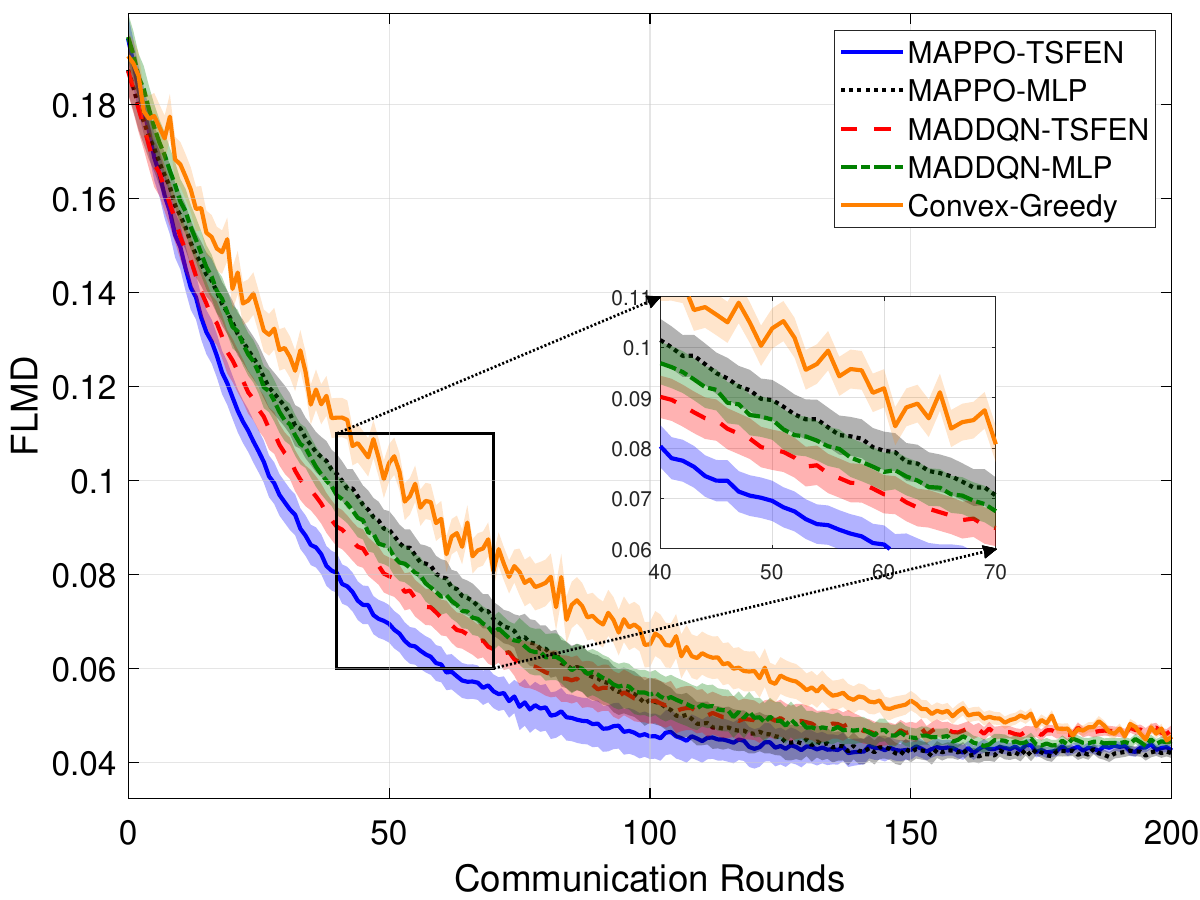}
    \caption{$K=6$}
\end{subfigure}
\caption{\textcolor{blue}{FLMD comparison between our proposed approach and baseline methods under different subchannel configurations.}}
\label{FLMD}
\end{figure*}

\begin{figure*}[htbp]
\centering
\begin{subfigure}[b]{0.3\textwidth}
    \includegraphics[width=\textwidth]{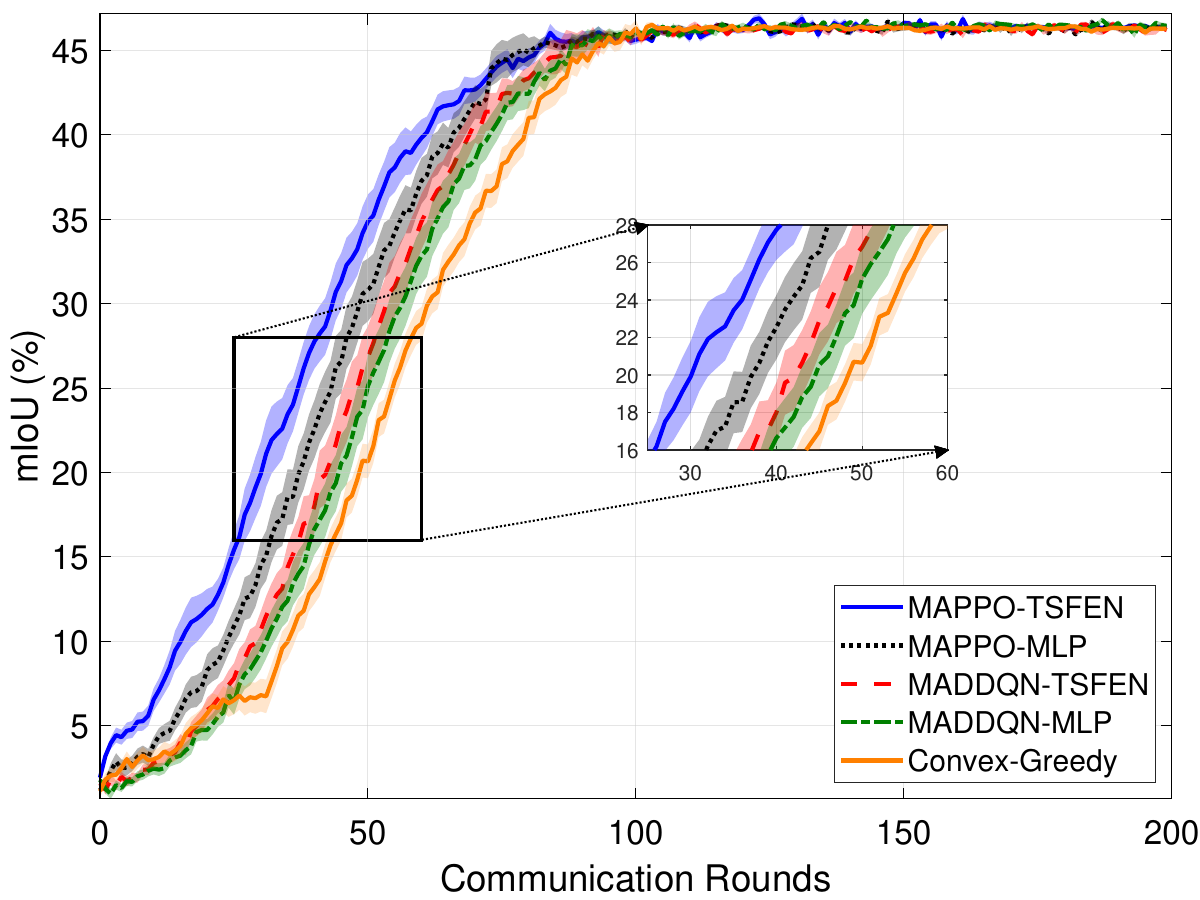}
    \caption{$K=2$}
\end{subfigure}
\hfill
\begin{subfigure}[b]{0.3\textwidth}
    \includegraphics[width=\textwidth]{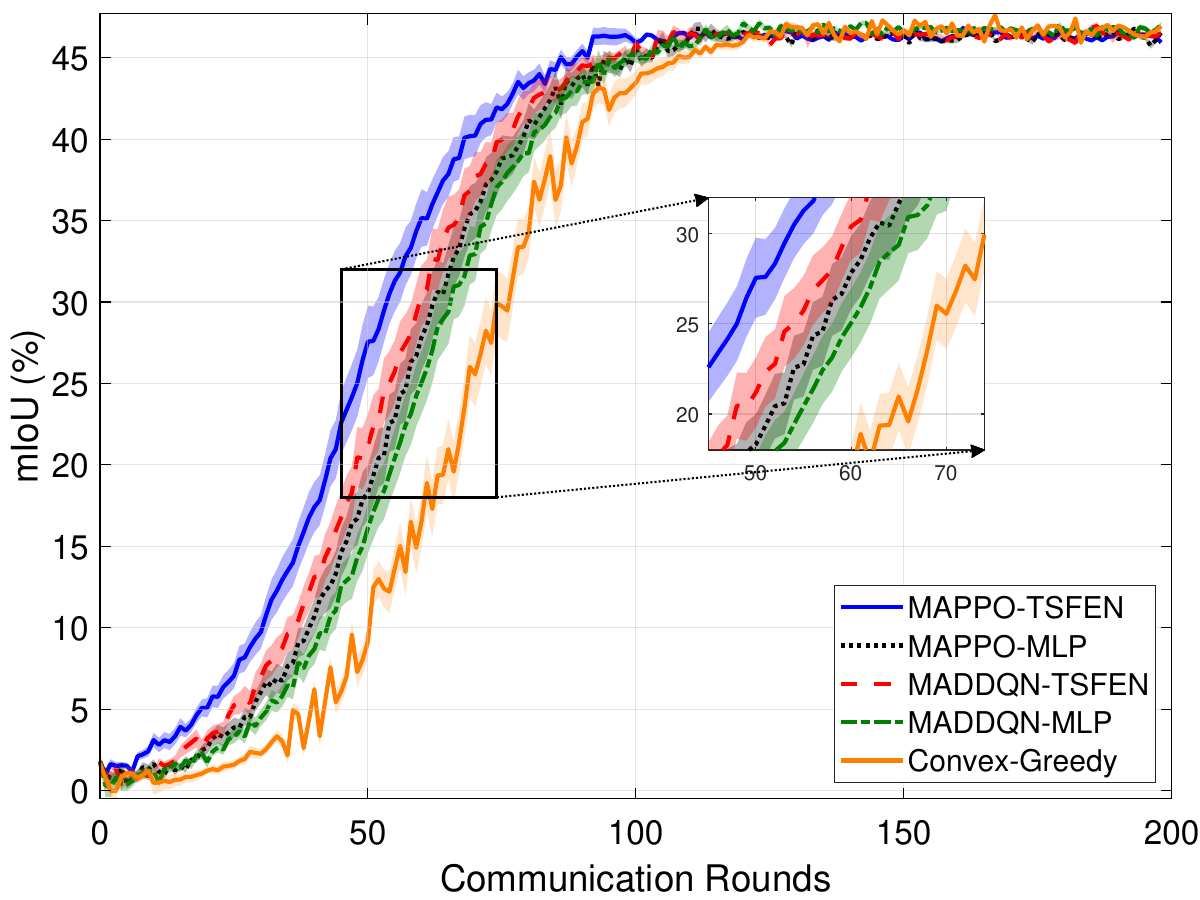}
    \caption{$K=4$}
\end{subfigure}
\hfill
\begin{subfigure}[b]{0.3\textwidth}
    \includegraphics[width=\textwidth]{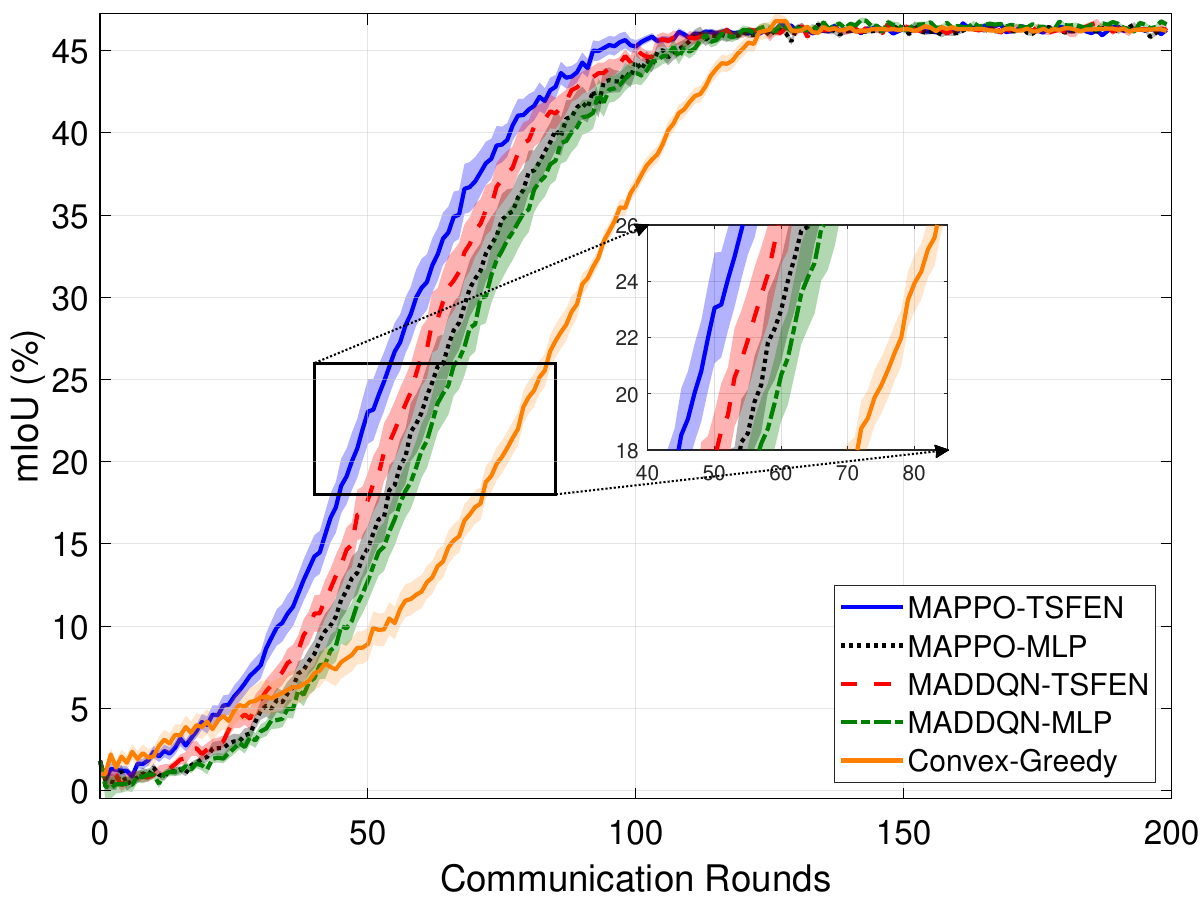}
    \caption{$K=6$}
\end{subfigure}
\caption{\textcolor{blue}{mIoU comparison on the AI4MARS dataset for our approach and baseline methods.}}
\label{mIoU}
\end{figure*}

\subsection{Performance Metrics}
To comprehensively evaluate our resource-aware control framework for WFL-empowered VP systems, we assess performance using multiple metrics:

\begin{enumerate}
    \item \textbf{Reward}: We track the cumulative reward during training, which combines AoI and FLMD optimization objectives. This metric reveals learning dynamics and convergence properties of our proposed MAPPO approach compared to baseline methods.
    
    \item \textbf{Sum AoI}: This metric quantifies the total freshness of information across all vehicles in the VP system. Lower Sum AoI indicates more timely model updates and better real-time performance in dynamic vehicular environments.
    
    \item \textbf{FLMD}: We measure Federated Learning Model Drift to evaluate the consistency between local and global models. This metric directly impacts model security and learning efficiency as established in our theoretical analysis.
    
    \item \textbf{Mean IoU}: For terrain classification performance on the AI4MARS dataset, we use mean Intersection over Union (mIoU), which provides a balanced assessment of segmentation accuracy across all terrain classes, particularly important for imbalanced datasets.
\end{enumerate}

\subsection{Results}

\subsubsection{Reward Convergence Analysis}
Fig. \ref{Reward} illustrates the reward curves during training across different subchannel configurations. As observed, our proposed approach with MHSA network architecture consistently achieves higher rewards compared to baseline methods, with convergence occurring around 60-80 episodes. This demonstrates that our approach effectively captures the dynamic nature of vehicular platooning environments.

\textcolor{blue}{The performance gap between our MAPPO-MHSA and baseline methods widens as system complexity increases, highlighting the scalability advantage of our temporal sequence feature extraction and multi-head self-attention mechanisms in handling complex communication topologies. While baseline methods exhibit increasing convergence instability in larger configurations, our approach maintains stable convergence across all scenarios, confirming the superiority of our optimization technique for dynamic vehicular environments.}

\subsubsection{AoI Optimization}
Fig. \ref{AoI} presents the Sum AoI performance across different methods and \textcolor{blue}{platoon configurations. The results demonstrate that our RACE framework combined with MAPPO-MHSA vehicle selection consistently achieves the lowest Sum AoI compared to MADDQN variants and the Convex-Greedy baseline, significantly improving information freshness across all scenarios.}

\textcolor{blue}{Our approach maintains consistent AoI optimization performance across the entire spectrum from small to large platoons, achieving 15-25\% lower AoI values compared to state-of-the-art baselines. While baseline methods show degraded performance in larger configurations (particularly evident in the $N=30, K=6$ scenario), our MAPPO-MHSA approach demonstrates remarkable scalability by maintaining low AoI values even as system complexity increases, indicating that our sophisticated optimization becomes more valuable in complex scenarios where traditional heuristics struggle with coordination overhead.}

\subsubsection{FLMD Convergence}
Fig. \ref{FLMD} shows the FLMD convergence patterns for different approaches. \textcolor{blue}{The key advantage of our approach is 40-60\% faster convergence rate compared to baseline methods. Our MAPPO-TSFEN method achieves rapid and stable FLMD convergence within 40-60 rounds regardless of system scale, while baseline methods experience significantly slower convergence in larger platoons, requiring up to 100+ communication rounds in complex configurations.}

This accelerated convergence validates our theoretical analysis in Section \ref{ProFm}, where we established the connection between optimization objectives and FLMD reduction. \textcolor{blue}{Our adaptive probability masking mechanism proves increasingly effective in filtering problematic vehicles as platoon size increases, providing crucial scalability advantages for large platoon deployments where model drift can propagate quickly through the network.}

\subsubsection{Terrain Classification Performance}
Fig. \ref{mIoU} displays the mIoU curves on the AI4MARS dataset across different methods and \textcolor{blue}{scalability configurations. Our MAPPO-TSFEN approach demonstrates 20-30 rounds faster mIoU convergence compared to baseline methods, with the convergence speed advantage becoming more pronounced in complex scenarios.}

\textcolor{blue}{Across all configurations, our approach maintains consistent rapid learning rates while baseline methods require increasingly more communication rounds as system complexity grows. This scalability characteristic is essential for practical deployments where large platoons must adapt quickly to diverse terrain conditions.}

The results confirm that our method's focus on optimizing AoI and accelerating FLMD convergence translates directly to more efficient model training, without compromising final model accuracy. \textcolor{blue}{The performance gap is most noticeable during the steep learning phase, highlighting our method's superior learning efficiency in dynamic vehicular environments.}

\textcolor{blue}{Overall, experimental results across all scales demonstrate our MAPPO-TSFEN approach achieves 15-25\% lower AoI values, 40-60\% faster FLMD convergence, and 20-30 rounds faster mIoU convergence compared to state-of-the-art baselines, validating the superior scalability of our joint AoI-FLMD optimization framework for future smart transportation systems with larger vehicle networks.}

\section{Conclusion} \label{Conclusion}

In this paper, we addressed the challenge of achieving timely and accurate control in WFL-empowered VP systems operating in highly dynamic environments. To this end, we formulated a joint optimization problem that simultaneously considers AoI and FLMD to ensure the timeliness and accuracy of control. To solve the problem, we designed a \underline{\textbf{R}}esource-\underline{\textbf{A}}ware \underline{\textbf{C}}ontrol fram\underline{\textbf{E}}work (\textbf{RACE}). The framework consists of two stages: first, a Lagrangian dual decomposition method for resource configuration, and second, a multi-agent proximal policy optimization approach for vehicle selection. Based on our theoretical analysis of FLMD's impact on convergence performance, we further designed an adaptive probability masking mechanism to optimize device selection strategies. Experimental results on the AI4MARS dataset show that, compared to baseline methods, our RACE framework significantly optimizes information timeliness with up to 45\% improvement in AoI while accelerating learning convergence, enabling faster adaptation in terrain classification tasks.

\appendix

\section*{Appendix A: Proofs}

\subsection*{Proof of Lemma 3}
\begin{proof}
We define a random variable $X$ as:
\begin{equation}
X = \frac{1}{K}\sum_{n\in\mathcal{S}[t]}\zeta_n\nabla f_n(\omega[t]) - \frac{1}{N}\sum_{n\in\mathcal{N}}\zeta_n\nabla f_n(\omega[t]).
\end{equation}
Since $F_r(\omega[t]) = \frac{1}{N}\sum_{i=1}^N f_r(\omega[t])$, we have:
\begin{equation}
    \begin{split}
        &\mathbb{E}[\|e[t]\|^2] = \mathbb{E}[\|X\|^2] \\
        &= \mathbb{E}\left[\left\|\frac{1}{K}\sum_{n\in\mathcal{S}[t]}\zeta_n\nabla f_n(\omega[t]) - \nabla F(\omega[t], \mathcal{N})\right\|^2\right].
    \end{split}
\end{equation}
By analyzing the variance of the cluster sampling estimator with unequal cluster sizes and applying the Cauchy-Schwarz inequality, we obtain the result.
\end{proof}

\subsection*{Proof of Theorem 4}
\begin{proof}
According to the Lipschitz smoothness property in \textbf{Assumption 1}, we have:
\begin{multline}
\frac{1}{2L}\|\nabla F(\omega[t], \mathcal{N}) - \nabla F(\omega[t+1], \mathcal{N})\|^2 \leq \\
F(\omega[t+1], \mathcal{N}) - F(\omega[t], \mathcal{N}) + \\
[\nabla F(\omega[t], \mathcal{N})^{\mathrm{T}}(\omega[t] - \omega[t+1])] \\
\leq \frac{L}{2}\|\omega[t] - \omega[t+1]\|^2.
\end{multline}
From Eq. \eqref{eq:global model} and the definition of $e[t]$, we have:
\begin{equation}
\begin{split}
\omega[t] - \omega[t+1] &= \xi\nabla F(\omega[t], \mathcal{S}[t]) \\
&= \xi[\nabla F(\omega[t], \mathcal{N}) + e[t]].
\end{split}
\end{equation}
Substituting this into the previous inequality:
\begin{align*}
F&(\omega[t+1], \mathcal{N}) \leq F(\omega[t], \mathcal{N}) \\
&- \xi\nabla F(\omega[t], \mathcal{N})^{\mathrm{T}}[\nabla F(\omega[t], \mathcal{N}) + e[t]] \\
&+ \frac{\xi^2 L}{2}\|\nabla F(\omega[t], \mathcal{N}) + e[t]\|^2 \\
&= F(\omega[t], \mathcal{N}) - \xi\|\nabla F(\omega[t], \mathcal{N})\|^2 \\
&- \xi\nabla F(\omega[t], \mathcal{N})^{\mathrm{T}} e[t] + \frac{\xi^2 L}{2}[\|\nabla F(\omega[t], \mathcal{N})\|^2 \\
&+ \|e[t]\|^2 + 2\nabla F(\omega[t], \mathcal{N})^{\mathrm{T}} e[t]].
\end{align*}
With learning rate $\xi = \frac{1}{L}$, this simplifies to:
\begin{equation}
\begin{split}
F(\omega[t+1], \mathcal{N}) \leq & F(\omega[t], \mathcal{N}) - \frac{1}{2L}\|\nabla F(\omega[t], \mathcal{N})\|^2 \\
&+ \frac{1}{2L}\|e[t]\|^2.
\end{split}
\end{equation}
By subtracting $F(\omega^*)$, taking expectations, and applying \textbf{Assumption 2}:
\begin{align*}
\mathbb{E}&[F(\omega[t+1], \mathcal{N}) - F(\omega^*)] \leq \\
&\mathbb{E}[F(\omega[t], \mathcal{N}) - F(\omega^*)] - \frac{1}{2L}\mathbb{E}[\|\nabla F(\omega[t], \mathcal{N})\|^2] \\
&+ \frac{1}{2L}\mathbb{E}[\|e[t]\|^2] \\
&\leq \left(1 - \frac{\mu}{L}\right)\mathbb{E}[F(\omega[t], \mathcal{N}) - F(\omega^*)] + \frac{1}{2L}\mathbb{E}[\|e[t]\|^2].
\end{align*}
Applying this recursively and summing over $t$, we obtain the bound.
\end{proof}

\subsection*{Proof of Theorem 5}

\begin{proof}
From Lemma 3, we can upper bound the deviation term by taking the maximum over all vehicles:
\begin{equation}
\mathbb{E}[\|e[t]\|^2] \leq \frac{(1-\frac{K}{N})\gamma^2}{K(N-1)\bar{\zeta}^2} \leq \frac{(1-\frac{K}{N})\gamma^2}{K\bar{\zeta}^2}.
\end{equation}

Starting from the general convergence bound, we have:
\begin{multline}
\mathbb{E}[F(\omega[t+1], \mathcal{N}) - F(\omega^*)] \leq \\
\left(1 - \frac{\mu}{L}\right)^t\mathbb{E}[F(\omega[1], \mathcal{N}) - F(\omega^*)] \\
+ \frac{1}{2L}\sum_{i=1}^{t}\left(1 - \frac{\mu}{L}\right)^{t-i}\mathbb{E}[\|e[i]\|^2].
\end{multline}

Substituting our bound on $\mathbb{E}[\|e[i]\|^2]$:
\begin{multline}
\mathbb{E}[F(\omega[t+1], \mathcal{N}) - F(\omega^*)] \leq \\
\left(1 - \frac{\mu}{L}\right)^t\mathbb{E}[F(\omega[1], \mathcal{N}) - F(\omega^*)] \\
+ \frac{1}{2L}\sum_{i=1}^{t}\left(1 - \frac{\mu}{L}\right)^{t-i}\frac{(1-\frac{K}{N})\gamma^2}{K\bar{\zeta}^2}.
\end{multline}

Factoring out constants from the summation:
\begin{multline}
= \left(1 - \frac{\mu}{L}\right)^t\mathbb{E}[F(\omega[1], \mathcal{N}) - F(\omega^*)] \\
+ \frac{(1-\frac{K}{N})\gamma^2}{2LK\bar{\zeta}^2}\sum_{i=1}^{t}\left(1 - \frac{\mu}{L}\right)^{t-i},
\end{multline}
which completes the proof.
\end{proof}

\subsection*{Proof of Theorem 7}

\begin{proof}
Let $\mathcal{M}_{a}[t] = \{n \in \mathcal{N} : \Theta_n[t] \leq \Lambda_{\Theta}[t]\}$ (eligible vehicles with adaptive threshold) and $\mathcal{M}_{f}[t] = \{n \in \mathcal{N} : \Theta_n[t] \leq \Lambda_{\Theta}^{\min}\}$ (eligible vehicles with fixed threshold). Since $\Lambda_{\Theta}[t] \geq \Lambda_{\Theta}^{\min}$, we have $\mathcal{M}_{f}[t] \subseteq \mathcal{M}_{a}[t]$.

From Lemma 3, the deviation bound for the adaptive case becomes:
\begin{multline}
\mathbb{E}[\|e_{a}[t]\|^2] \leq \\
\frac{(1-\frac{K}{N})\sum_{n\in\mathcal{M}_{a}[t]}\zeta_n^2\|\nabla f_n(\omega[t])-\nabla F(\omega[t], \mathcal{N})\|^2}{K(N-1)\left(\frac{1}{N}\sum_{n\in\mathcal{N}}\zeta_n\right)^2}
\end{multline}

\begin{multline}
\leq \frac{|\mathcal{M}_{a}[t]|}{|\mathcal{M}_{f}[t]|} \cdot \\
\frac{(1-\frac{K}{N})\sum_{n\in\mathcal{M}_{f}[t]}\zeta_n^2\|\nabla f_n(\omega[t])-\nabla F(\omega[t], \mathcal{N})\|^2}{K(N-1)\left(\frac{1}{N}\sum_{n\in\mathcal{N}}\zeta_n\right)^2} \\
= \rho[t] \cdot \mathbb{E}[\|e[t]\|^2],
\end{multline}
where the second inequality uses the fact that vehicles in $\mathcal{M}_{a}[t] \setminus \mathcal{M}_{f}[t]$ have bounded FLMD values.

From the general convergence bound, we have:
\begin{multline}
\mathbb{E}[F(\omega[t+1], \mathcal{N}) - F(\omega^*)] \leq \\
\left(1 - \frac{\mu}{L}\right)^t\mathbb{E}[F(\omega[1], \mathcal{N}) - F(\omega^*)] \\
+ \frac{1}{2L}\sum_{i=1}^{t}\left(1 - \frac{\mu}{L}\right)^{t-i}\mathbb{E}[\|e[i]\|^2].
\end{multline}

For the adaptive threshold case, we replace $\mathbb{E}[\|e[i]\|^2]$ with $\mathbb{E}[\|e_a[i]\|^2]$ in the deviation term. Using our established bound $\mathbb{E}[\|e_a[i]\|^2] \leq \rho[i] \cdot \mathbb{E}[\|e[i]\|^2]$, the convergence analysis applies directly with the adaptive threshold mechanism, yielding the desired result.
\end{proof}

\subsection*{Proof of Theorem 8}

\begin{proof}
Under local smoothness, the Lipschitz condition holds within $\mathcal{B}_R(\omega^*)$. Using concentration inequalities, we can show that with probability $1-\delta$, the iterates remain within the smoothness region, guaranteeing convergence.

Let $X_t = \|\omega[t] - \omega^*\|$ denote the distance from the optimal point at iteration $t$. From the algorithm update rule and local smoothness property:
\begin{equation}
X_{t+1} \leq X_t - \xi\|\nabla F(\omega[t], \mathcal{N})\|^2/L + \xi\|e[t]\|.
\end{equation}

By Azuma's inequality, since $\|e[t]\|$ has bounded differences, we have:
\begin{equation}
\mathbb{P}\left(\sum_{t=1}^T \xi\|e[t]\| \geq \epsilon/2\right) \leq \exp\left(-\frac{(\epsilon/2)^2}{2T\xi^2\sigma^2}\right),
\end{equation}
where $\sigma^2$ is the variance bound of $\|e[t]\|$.

Setting $\xi \leq \frac{\epsilon}{2LT\sqrt{\log(1/\delta)}}$ and using $\sigma^2 \leq L^2$, we get:
\begin{equation}
\mathbb{P}\left(\sum_{t=1}^T \xi\|e[t]\| \geq \epsilon/2\right) \leq \delta/2.
\end{equation}

Similarly, the gradient descent progress ensures:
\begin{equation}
\mathbb{P}\left(\sum_{t=1}^T \xi\|\nabla F(\omega[t], \mathcal{N})\|^2/L \geq \epsilon/2\right) \leq \delta/2.
\end{equation}

By union bound, with probability $1-\delta$:
\begin{equation}
X_T \leq X_0 - \epsilon/2 + \epsilon/2 = R - \epsilon \leq R,
\end{equation}
ensuring the iterates remain within the smoothness region.
\end{proof}

\subsection*{Proof of Theorem 9}

\begin{proof}
Following standard analysis for non-convex optimization, we telescope the smoothness inequality and apply Jensen's inequality to obtain the convergence rate to stationary points.

From the smoothness property (Assumption 1):
\begin{multline}
F(\omega[t+1], \mathcal{N}) \leq F(\omega[t], \mathcal{N}) \\
+ \nabla F(\omega[t], \mathcal{N})^T(\omega[t+1] - \omega[t]) \\
+ \frac{L}{2}\|\omega[t+1] - \omega[t]\|^2.
\end{multline}

Substituting $\omega[t+1] - \omega[t] = -\xi(\nabla F(\omega[t], \mathcal{N}) + e[t])$:
\begin{multline}
F(\omega[t+1], \mathcal{N}) \leq F(\omega[t], \mathcal{N}) - \xi\|\nabla F(\omega[t], \mathcal{N})\|^2 \\
- \xi\nabla F(\omega[t], \mathcal{N})^Te[t] \\
+ \frac{\xi^2L}{2}\|\nabla F(\omega[t], \mathcal{N}) + e[t]\|^2.
\end{multline}

Expanding the last term and using $\xi = 1/L$:
\begin{multline}
F(\omega[t+1], \mathcal{N}) \leq F(\omega[t], \mathcal{N}) \\
- \frac{1}{2L}\|\nabla F(\omega[t], \mathcal{N})\|^2 + \frac{1}{2L}\|e[t]\|^2.
\end{multline}

Rearranging and taking expectations:
\begin{multline}
\mathbb{E}[\|\nabla F(\omega[t], \mathcal{N})\|^2] \leq \\
2L(\mathbb{E}[F(\omega[t], \mathcal{N})] - \mathbb{E}[F(\omega[t+1], \mathcal{N})]) \\
+ \mathbb{E}[\|e[t]\|^2].
\end{multline}

Summing over $t = 1, \ldots, T$:
\begin{multline}
\sum_{t=1}^T \mathbb{E}[\|\nabla F(\omega[t], \mathcal{N})\|^2] \leq \\
2L(\mathbb{E}[F(\omega[1], \mathcal{N})] - \mathbb{E}[F(\omega[T+1], \mathcal{N})]) \\
+ \sum_{t=1}^T \mathbb{E}[\|e[t]\|^2].
\end{multline}

Since $F(\omega[T+1], \mathcal{N}) \geq F^*$:
\begin{equation}
\sum_{t=1}^T \mathbb{E}[\|\nabla F(\omega[t], \mathcal{N})\|^2] \leq 2L(F(\omega[0]) - F^*) + \sum_{t=1}^T \mathbb{E}[\|e[t]\|^2].
\end{equation}

Dividing by $T$ and taking the minimum:
\begin{multline}
\min_{t=1,\ldots,T} \mathbb{E}[\|\nabla F(\omega[t], \mathcal{N})\|^2] \\
\leq \frac{1}{T}\sum_{t=1}^T \mathbb{E}[\|\nabla F(\omega[t], \mathcal{N})\|^2] \\
\leq \frac{2L(F(\omega[0]) - F^*) + \sum_{t=1}^T \mathbb{E}[\|e[t]\|^2]}{T},
\end{multline}
which completes the proof.
\end{proof}

\subsection*{Proof of Theorem 10}
\begin{proof}
From the definition of FLMD and Eq. (4):
\begin{align*}
\Theta_n[t] &= \frac{\|\omega_n[t] - \omega[t]\|}{\|\omega[t]\|} \\
&= \frac{\left\|\omega[t] - \frac{\xi}{\zeta_n}\sum_{i=1}^{\zeta_n}\nabla\ell(\omega[t]; \mathbf{x}_{n,i}, y_{n,i}) - \omega[t]\right\|}{\|\omega[t]\|} \\
&= \frac{\left\|\frac{\xi}{\zeta_n}\sum_{i=1}^{\zeta_n}\nabla\ell(\omega[t]; \mathbf{x}_{n,i}, y_{n,i})\right\|}{\|\omega[t]\|} \\
&= \frac{\xi\|\nabla f_n(\omega[t])\|}{\|\omega[t]\|}.
\end{align*}
Using the Lipschitz continuity of gradients:
\begin{align*}
\|\nabla f_n(\omega[t])\| &\leq \|\nabla f_n(\omega[t]) - \nabla f_n(\omega^*)\| + \|\nabla f_n(\omega^*)\| \\
&\leq L\|\omega[t] - \omega^*\| + \|\nabla f_n(\omega^*)\|.
\end{align*}
Therefore:
\begin{equation}
\Theta_n[t] \leq \frac{\xi(L\|\omega[t] - \omega^*\| + \|\nabla f_n(\omega^*)\|)}{\|\omega[t]\|}.
\end{equation}
\end{proof}

\subsection*{Proof of Theorem 17}
\begin{proof}
Under high SNR, we can approximate $2^{\frac{D_n[t]}{\delta_n^{\mathrm{tx}} B}} - 1 \approx 2^{\frac{D_n[t]}{\delta_n^{\mathrm{tx}} B}}$. The energy constraint becomes:
\begin{equation}
\kappa \mu \zeta_n (\chi_n^* G_n)^2 + \delta_n^{\mathrm{tx}} \frac{2^{\frac{D_n[t]}{\delta_n^{\mathrm{tx}} B}}}{|h_n|^2} = e_n^{\max}.
\end{equation}

Solving for $2^{\frac{D_n[t]}{\delta_n^{\mathrm{tx}} B}}$:
\begin{equation}
2^{\frac{D_n[t]}{\delta_n^{\mathrm{tx}} B}} = \frac{(e_n^{\max} - \kappa \mu \zeta_n (\chi_n^* G_n)^2)|h_n|^2}{\delta_n^{\mathrm{tx}}}.
\end{equation}

Taking logarithms:
\begin{equation}
\frac{D_n[t]}{\delta_n^{\mathrm{tx}} B} = \log_2\left(\frac{(e_n^{\max} - \kappa \mu \zeta_n (\chi_n^* G_n)^2)|h_n|^2}{\delta_n^{\mathrm{tx}}}\right).
\end{equation}

Rearranging to solve for $\delta_n^{\mathrm{tx}}$:
\begin{equation}
\delta_n^{\mathrm{tx}} = \frac{D_n[t]}{B \log_2\left(1 + \frac{e_n^{\max} |h_n|^2}{\kappa \mu \zeta_n (\chi_n^* G_n)^2}\right)},
\end{equation}

which completes the proof.
\end{proof}

\section*{Appendix B: Complete Implementation Specifications}

Our TSFEN architecture consists of three main components with the following specifications:
\begin{itemize}
\item Multi-Head Self-Attention: $d_{\text{model}} = 64$, $h = 8$ heads, input shape $(B, M=5, N=20, 64)$
\item LSTM Component: $d_{\text{hidden}} = 128$, bidirectional processing
\item Actor-Critic Networks: Linear layers $(128 \to 128 \to 20)$ with ReLU activation
\end{itemize}

The training configuration for our MAPPO implementation includes:
\begin{itemize}
\item MAPPO: $\beta = 10^{-4}$, $\gamma = 0.98$, $\lambda = 0.95$, $\epsilon = 0.2$
\item Batch size: 32 trajectories, Episodes per update: 10, Total episodes: 500
\item State representation: $s_{k,m}[t] = (\Theta_n[t], |h_n|^2, \Delta_n[t])$ for $n \in \mathcal{N}$
\end{itemize}

For the Lagrangian dual decomposition implementation, we specify:
\begin{itemize}
\item Numerical solver: scipy.optimize.brentq with tolerance $\epsilon = 10^{-6}$
\item Maximum iterations: 100, Search interval: $[\frac{D_n[t]}{B \log_2(1 + P_n |h_n|^2)}, \delta_{\max}]$
\end{itemize}

Our implementation environment comprises:
\begin{itemize}
\item Hardware: NVIDIA RTX 5070Ti with CUDA 12.8
\item Software: PyTorch 1.12.0, Python 3.10, scipy.optimize for numerical solving
\item Memory requirement: $\sim 2.3$ GB for $N=20, K=6$ configuration
\end{itemize}

Critical implementation details include:
\begin{itemize}
\item Numerical stability: Action probabilities clamped to $[10^{-7}, 1.0]$ and renormalized
\item FLMD computation: $\Theta_n[t] = \frac{\xi \|\nabla f_n(\omega[t])\|}{\|\omega[t]\|}$ after each local training step
\item Adaptive masking: Element-wise multiplication with mask vector before softmax
\item Vehicle dynamics: Updated using Eqs. (10)-(11) with $\tau = 1$s per communication round
\end{itemize}

\section*{Appendix C: Edge Device Feasibility Analysis}

Our framework is designed for realistic vehicular edge computing platforms:
\begin{itemize}
\item Mobile terminals: Raspberry Pi 5-based edge devices with ARM Cortex-A76 quad-core processor (2.4 GHz), 8 GB LPDDR4X-4267 SDRAM, and 64 GB microSD storage
\item Communication subsystem: NI USRP-2900 units (1 Tx/1 Rx, 56 MHz bandwidth) for vehicular communication, along with ROS2-compatible Lidar and 3D depth cameras
\item Platoon leader: NI USRP-2955 (4 Rx channels, 80 MHz bandwidth), Intel i7 processor with 16 GB RAM for model aggregation, and Ethernet/MXIe connectivity
\end{itemize}

The computational complexity analysis reveals manageable requirements for each framework component:
\begin{itemize}
\item MHSA module: Complexity $O(M \cdot N^2 \cdot d_{\text{model}})$ where $M = 5$, $N = 20$, and $d_{\text{model}} = 64$, requiring approximately 40 KB for attention matrices with projected computation time of 2.3 ms on ARM Cortex-A76
\item LSTM component: Complexity $O(M \cdot d_{\text{lstm}}^2)$ where $d_{\text{lstm}} = 256$, needs around 512 KB for hidden states and completes processing in 1.8 ms per sequence
\item Actor-critic networks: Linear transformations with less than 100K total parameters, requiring 400 KB memory and under 0.5 ms inference time
\end{itemize}

For practical deployment, our terrain classification model employs several optimizations:
\begin{itemize}
\item Backbone optimization: MobileNetV2-based feature extraction instead of full ResNet101, reducing parameters from 59M to 2.5M (95\% reduction) with approximately 10 MB memory footprint and 45 ms inference time
\item Quantization: 16-bit precision provides 50\% memory reduction compared to 32-bit floating point, cutting transmission overhead from 200+ MB to 2.5 MB per model update
\item Layer-wise updates: Only task-specific layers participate in federated training, with 2.1M trainable parameters requiring 12 ms for gradient computation per local training iteration
\end{itemize}

The Lagrangian dual decomposition solver implementation details are:
\begin{itemize}
\item Solver method: Modified Brent's method with tolerance $\epsilon = 10^{-6}$
\item Computational complexity: $O(\log(\Delta/\epsilon))$, typically requiring 15 iterations per vehicle
\item Execution time: Under 5 ms per vehicle with less than 1 KB memory requirement for solver state variables
\end{itemize}

System-level performance analysis shows encouraging results:
\begin{itemize}
\item Memory usage: 1.2 MB for TSFEN model storage, 10 MB for local terrain classification, 8 MB for MAPPO training buffers, and 200 MB for system overhead, totaling approximately 220 MB (2.7\% of available 8 GB memory)
\item Runtime latency: 45 ms for model inference, 4.1 ms for TSFEN feature extraction, 5 ms for resource allocation computation, and under 1 ms for FLMD calculation, resulting in total per-round latency under 60 ms (well below 100 ms real-time constraint)
\item Energy consumption: 2.1 J computational energy, 1.8 J communication energy, and 0.3 J idle processing per round, totaling 4.2 J per FL round (less than 2\% battery life impact per hour for typical 5000 mAh vehicular batteries)
\end{itemize}

Our scalability analysis demonstrates appropriate scaling characteristics:
\begin{itemize}
\item Vehicle scaling: From $N = 10$ to $N = 30$ increases memory linearly ($\sim 3.0 \times$) and computation time by $\sim 4.2 \times$
\item Subchannel scaling: From $K = 2$ to $K = 6$ has negligible impact on individual vehicle computation
\item Model complexity: Linear relationship between model size and resource requirements
\end{itemize}



\bibliographystyle{IEEEtran}
                                                               
\bibliography{Ref}

\begin{IEEEbiography}[{\includegraphics[width=1in,height=1.25in,clip,keepaspectratio]{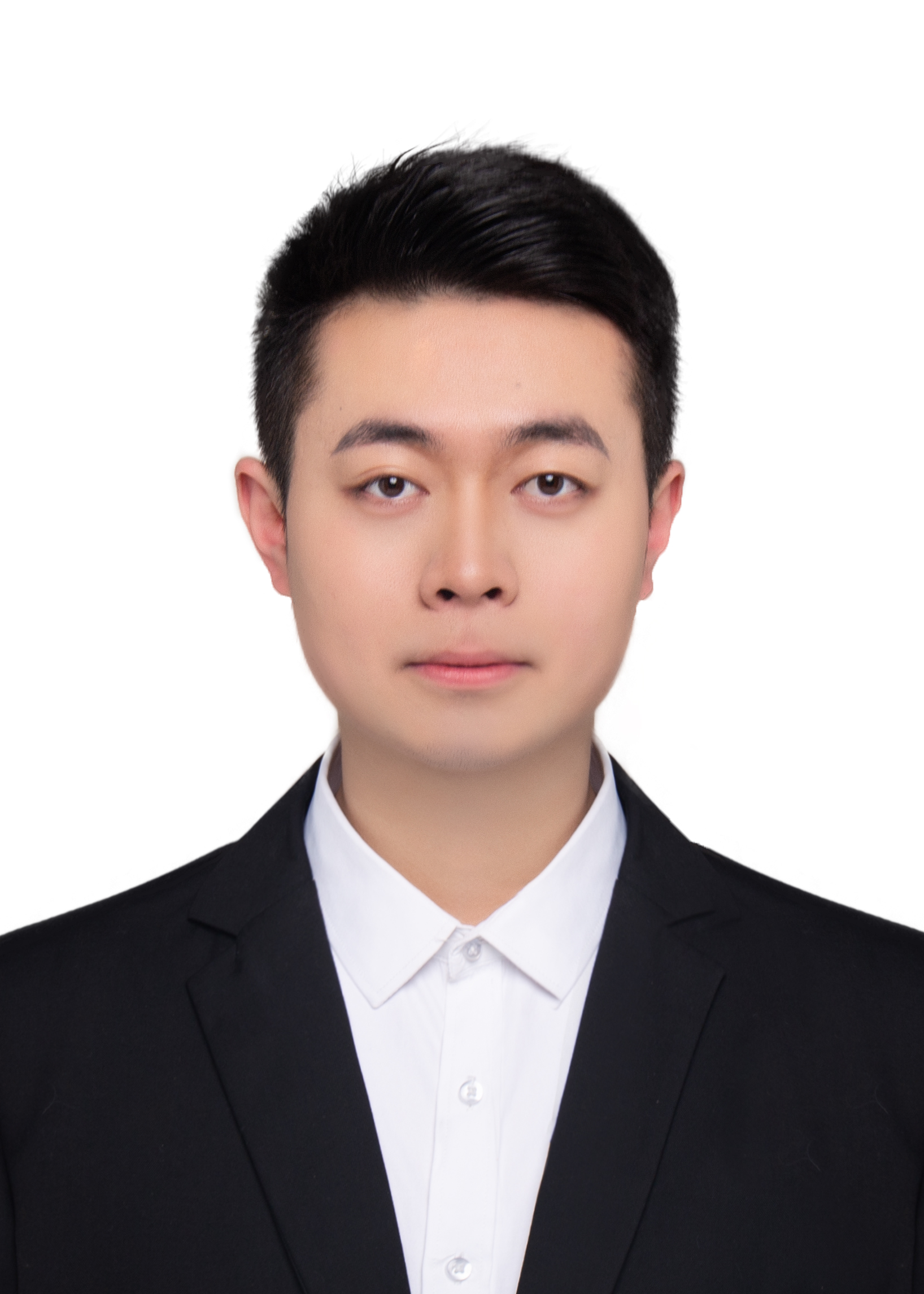}}]{Beining Wu} (Member, IEEE) received his BS degree in mathematics and applied mathematics from Anhui Normal University, Wuhu, China in 2024. He is currently pursuing a Ph.D. degree in Computer Science at South Dakota State University (SDSU), Brookings, United States. His research interests include wireless communications, UAV networks, and reinforcement learning.
\end{IEEEbiography}

\begin{IEEEbiography}[{\includegraphics[width=1in,height=1.25in,clip,keepaspectratio]{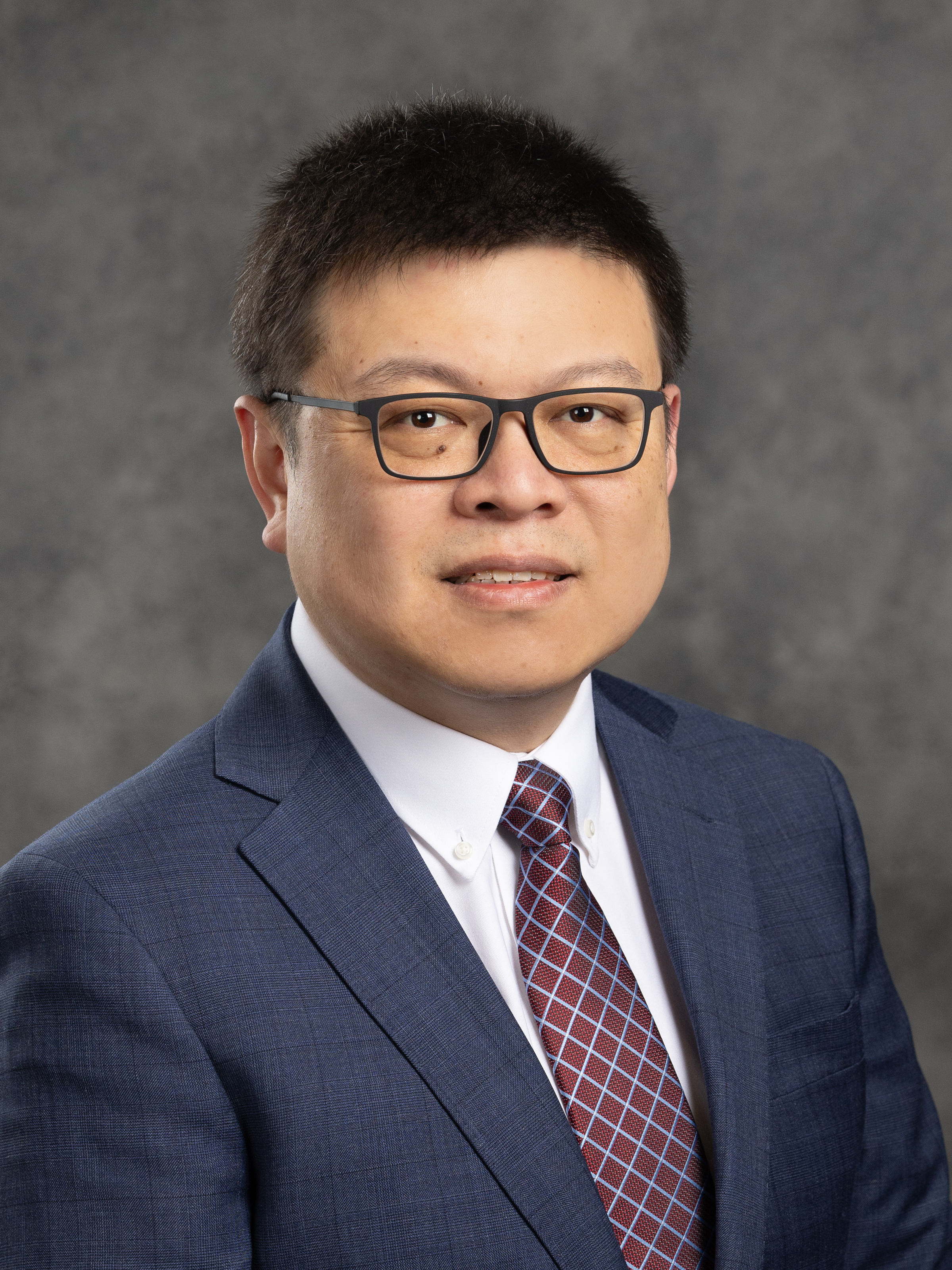}}]{Jun Huang} (M’12-SM’16) received a Ph.D. degree (with honors) from the Institute of Network Technology, Beijing University of Posts and Telecommunications, China, in 2012. He is now an Assistant Professor in the Department of Electrical Engineering and Computer Science (EECS) at South Dakota State University. Before that, he was a non-tenure track faculty member at Baylor University. He held a full professor appointment at Northwestern Polytechnical University and Chongqing University of Posts and Telecommunications in China from 2015 to 2021. Dr. Huang was a Visiting Scholar at the University of British Columbia, a Research Fellow at the South Dakota School of Mines \& Technology and the University of Texas at Dallas, and a Guest Professor at the National Institute of Standards and Technology. He was the recipient of the Outstanding Research Award (Tier I) from CQUPT in 2019, the Best Paper Award from EAI Mobimedia in 2019, Outstanding Service Award from ACM RACS in 2017, 2018, and 2019, Best Paper Nomination from ACM SAC in 2014, and Best Paper Award from AsiaFI 2011. He is the Technical Editor of ACM SIGAPP Applied Computing Review and an Associate Editor of Elsevier Digital Communications and Networks and ICT Express. He guest-edited several special issues in IEEE/ACM journals. He also chaired and co-chaired multiple conferences in the communications and networking areas and organized numerous workshops at major IEEE and ACM events. He is a Senior Member of the IEEE.
\end{IEEEbiography}

\begin{IEEEbiography}[{\includegraphics[width=1in,height=1.25in,clip,keepaspectratio]{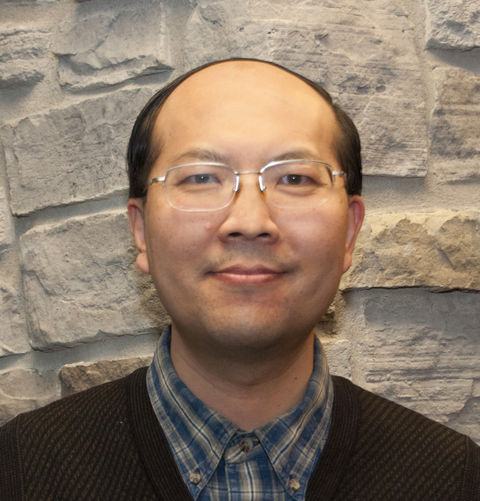}}]{Qiang Duan} (Senior Member, IEEE) is a Professor of Information Sciences and Technology at Pennsylvania State University Abington College. His general research interests include computer networking, distributed systems, and artificial intelligence, with recent research focusing on network virtualization and softwarization, network-edge-cloud convergence, federated and split learning, and ubiquitous intelligence in future Internet. Prof. Duan has published four monographs, six book chapters, and 120+ refereed journal articles and conference papers. He has served on the editorial boards as an editor/associate editor for multiple research journals and has been involved in organizing numerous international conferences as a TPC member and track/session chair. Prof. Duan received his Ph.D. in Electrical Engineering from the University of Mississippi in 2003. He is a Senior Member of the IEEE.
\end{IEEEbiography}

\begin{IEEEbiography}[{\includegraphics[width=1in,height=1.25in,clip,keepaspectratio]{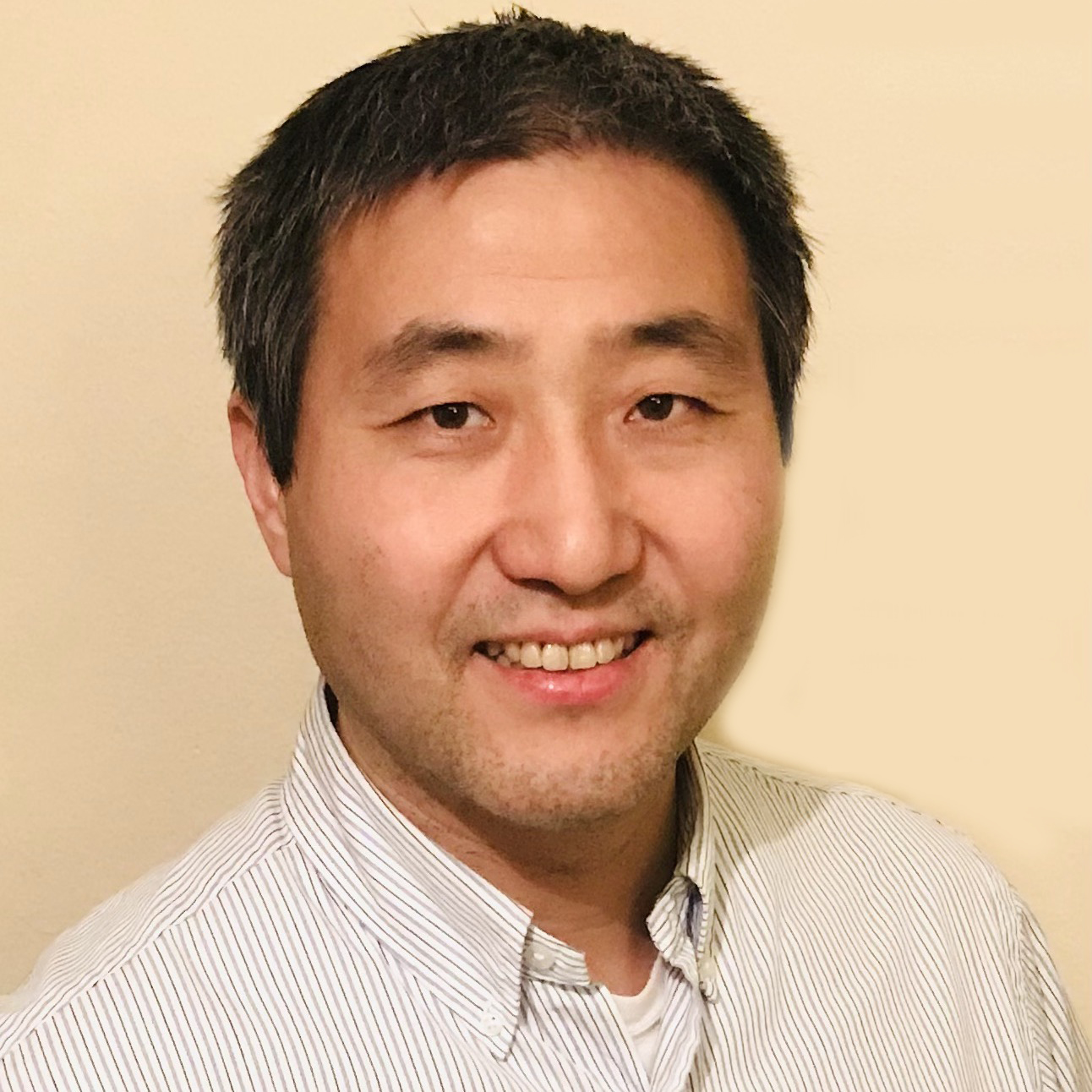}}]{Liang (Leon) Dong} (Senior Member, IEEE) received his B.S. degree in applied physics with a minor in computer engineering from Shanghai Jiao Tong University, China, in 1996, and his M.S. and Ph.D. degrees in electrical and computer engineering from the University of Texas at Austin in 1998 and 2002, respectively. Since 2011, he has been with Baylor University, where he is currently an Associate Professor of Electrical and Computer Engineering. His research interests include digital signal processing, wireless communications and networking, cyber-physical systems and security, and AI/ML applications in signal processing and communications. Dr. Dong's work has been supported by NSF, DoD, NASA, and industry partners such as L3Harris, Intel, and ExxonMobil. He has extensive industry experience in smart antenna communications systems and wireless networking technologies. Previously, he held academic positions at Western Michigan University and was a Visiting Researcher at Stanford University. Dr. Dong is a Senior Member of the Institute of Electrical and Electronics Engineers (IEEE) and a Member of the American Physical Society (APS).
\end{IEEEbiography}

\begin{IEEEbiography}[{\includegraphics[width=1in,height=1.25in,clip,keepaspectratio]{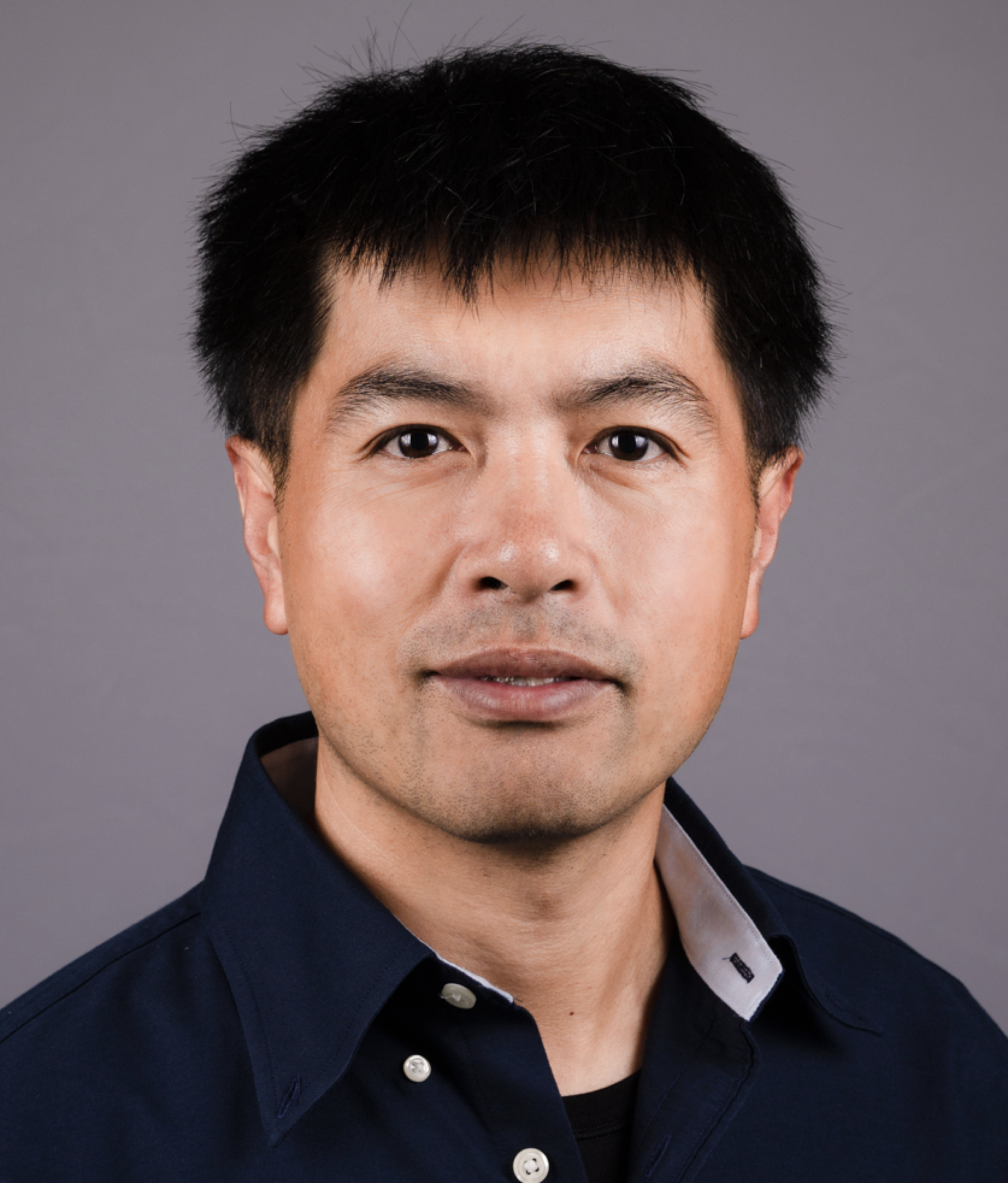}}]{Zhipeng Cai} (Fellow IEEE) Dr. Zhipeng Cai received his PhD and M.S. degrees in the Department of Computing Science at University of Alberta. He is currently a Professor in the Department of Computer Science at Georgia State University, and also an Affiliate Professor in the Department of Computer Information System at Robinson College of Business, as well as a Director at INSPIRE Center. Dr. Cai also leads the Innovative Computing and Networking (ICN) group. Dr. Cai's research expertise lies in Resource Management and Scheduling, High Performance Computing, Cyber-Security, Privacy, Networking, and Big Data. His research has received funding from multiple academic and industrial sponsors, including the National Science Foundation and the U.S. Department of State, and has resulted in over 100 publications in top journals and conferences, with more than 18,000 citations, including over 100 IEEE/ACM Transactions papers. He is listed as one of The World's Top 2\% Scientists (2020 - 2024, published by Stanford University). Additionally, ScholarGPS ranks him in the top 0.05\% of all scholars globally, acknowledging his significant scholarly contributions. Dr. Cai holds the esteemed position of Editor-in-Chief for Elsevier High-Confidence Computing Journal. He also serves as an editor for several prestigious journals, including IEEE TKDE, TVT, TWC, and TCSS. Moreover, Dr. Cai serves as a Steering Committee Co-Chair for WASA and is a Steering Committee member for COCOON, IPCCC, and ISBRA. He has also chaired numerous international conferences, including ICDCS, SocialCom, and ISBRA. 
Dr. Cai is the recipient of an NSF CAREER Award and is a Fellow of IEEE.

\end{IEEEbiography}

\end{document}